\documentclass[12pt,a4paper]{article}%
\pdfoutput = 1
\usepackage{amsfonts}
\usepackage{amssymb}
\usepackage[centertags]{amsmath}
\usepackage{graphicx}
\usepackage{lscape}
\usepackage[left=2.5cm, right=2.0cm, top=2.0cm, bottom = 2.5cm]{geometry}%
\providecommand{\U}[1]{\protect\rule{.1in}{.1in}}
\newtheorem{theorem}{Theorem}

\newtheorem{algorithm}[theorem]{Algorithm}

\ifx\pdfoutput\relax\let\pdfoutput=\undefined\fi
\newcount\msipdfoutput
\ifx\pdfoutput\undefined\else
\ifcase\pdfoutput\else
\ifx\paperwidth\undefined\else
\ifdim\paperheight=0pt\relax\else\pdfpageheight\paperheight\fi
\ifdim\paperwidth=0pt\relax\else\pdfpagewidth\paperwidth\fi
\fi\fi\fi
\begin{document}

\title{Inference on Self-Exciting Jumps in Prices and Volatility using High Frequency
Measures\thanks{The authors would like to thank three anonymous referees and a
co-editor for very detailed and constructive comments on an earlier draft of
the paper. We also thank Yacine A\"{\i}t-Sahalia, John Maheu, Eric Renault,
George Tauchen, Victor Todorov and Herman van Dijk for very constructive
comments at various stages in the development of the paper, plus the
participants at the Society of Financial Econometrics Annual Conference, 2014,
the International Association for Applied Econometrics Conference, 2014, and
the Econometric Society Australasian Meetings, 2014. The research has been
supported by Australian Research Council Discovery Grant DP150101728.} }
\author{Worapree Maneesoonthorn\thanks{Melbourne Business School, The University of
Melbourne. }, Catherine S. Forbes\thanks{Department of Econometrics and
Business Statistics, Monash University.} and Gael M.
Martin\thanks{Corresponding author: gael.martin@monash.edu. Department of
Econometrics and Business Statistics, Monash University.}}
\maketitle

\begin{abstract}
Dynamic jumps in the price and volatility of an asset are modelled using a
joint Hawkes process in conjunction with a bivariate jump diffusion. A state
space representation is used to link observed returns, plus nonparametric
measures of integrated volatility and price jumps, to the specified model
components; with Bayesian inference conducted using a Markov chain Monte Carlo
algorithm. An evaluation\textbf{ }of marginal likelihoods for the proposed
model relative to a large number of alternative models, including some that
have featured in the literature, is provided. An extensive empirical
investigation is undertaken using data on the S\&P500 market index over the
1996 to 2014 period, with substantial support for dynamic jump intensities -
including in terms of predictive accuracy - documented. \emph{\bigskip}


\end{abstract}

\baselineskip18pt\newpage

\section{Introduction}

Planning for unexpected and\textbf{\ }large movements in asset prices is
central to the management of financial risk. Key to this planning is the
ability to distinguish extreme price changes arising from a persistent shift
in the asset's underlying volatility from idiosyncratic movements that occur
due to random shocks in the market environment. Making this task more
difficult is the fact that volatility itself exhibits discontinuous behaviour
which, via the stylized occurrence of feedback from volatility to current and
future returns (e.g. Bollerslev, Sizova and Tauchen, 2012), has the potential
to cause seemingly discontinuous behaviour in the asset price. Moreover, it is
unclear whether the apparent clustering behaviour of asset price jumps during
times of market turbulence is evidence of dynamics in the jump
\textit{intensity} of either process (or both), or simply a result of the
propagation through time of (independent) variance jumps due to persistence in
the level of volatility.

Traditionally, parametric jump diffusion models have been used to capture the
discontinuous behaviour in prices and, potentially, in their underlying
volatility. Notable in this literature are the studies of Bates (2000) and Pan
(2002), which propose models that characterize the intensity of a jump in
price as proportional\ to the level of the underlying (diffusive) variance. In
these models, the (price) jump intensity will be high in periods with high
volatility and dependent over time as a consequence of the dynamic
specification adopted for volatility itself. Duffie, Pan and Singleton (2000),
on the other hand, introduce a model with both\ price and volatility jumps,
and where the contemporaneous occurrence of the two types of jumps (i.e. the
occurrence of `co-jumps') is imposed. Under this specification, large
fluctuations in price tend to occur in successive periods following a
(contemporaneous) jump in price and volatility, again due to persistence in
the volatility process.\ This impact is exacerbated by the fact that the
expected price\ jump size is assumed to be conditionally (positively)
dependent on the magnitude of the latent variance jump. Broadie, Chernov and
Johannes (2007) also specify co-jumps, but impose independence between the
sizes of the two different types of jump. Eraker, Johannes and Polson (2003),
Chernov, Gallant, Ghysels and Tauchen (2003) and Eraker (2004) use more
general specifications, in which both non-contemporaneous jumps and correlated
jump sizes are accommodated, although insignificant correlation between the
price and variance jump sizes is documented in all cases.

More recently, volatility and jump measures constructed from high frequency
data have been used to investigate price and variance jumps, including the
relationship between them. For example, the empirical findings of Todorov
and\ Tauchen (2011) indicate the presence of jumps in volatility, whilst those
of Jacod and Todorov (2010) provide evidence of both price and variance jumps,
with a certain proportion of those jumps occurring\ simultaneously for the
S\&P500 market index. Jacod, Kl\"{u}ppelberg and M\"{u}ller (2013)\textbf{
}use high frequency data to explore the correlation between (imposed) co-jumps
for several series, but fail to reject the null hypothesis of zero correlation
in the majority of cases considered.

As highlighted clearly by Bandi and Reno (2016), however, the use (or
otherwise) of option price data (and the associated risk premia
specifications) in past analyses, plus the very nature of the volatility
filter adopted (and data frequency exploited in the measurement of
volatility), is likely to have had an impact on conclusions drawn regarding
the joint evolution of a price and its variance, including discontinuities
therein; with such considerations possibly underlying the inconclusive results
recorded. We speculate that the rather restricted manner in which the
\textit{dynamics} in jumps have been modelled may also have played a role here.

With this background in mind, we propose a very general model for the joint
evolution of price and volatility in which both processes are permitted to
jump, co-jumps are possible (but not imposed), and both jump processes are
allowed to be dynamic. To this end, we adopt a bivariate Hawkes process
(Hawkes, 1971a,b) for the \textit{intensity} of price and variance
(accordingly volatility, defined as the square root of the variance) jumps,
with both jump processes being (potentially) self-exciting as a consequence;
that is, the intensity of each jump process is functionally dependent on the
realized past increments of that process. We allow the variance jump intensity
to depend on past price jumps, enabling\ extreme price movements to influence
the occurrence of extreme movements in volatility.\ Possible leverage effects
operating at the level of extreme price and volatility movements are also
accommodated via the modeling of the differential impacts of negative and
positive price jumps on the variance jump\textbf{\ }intensity.

A multivariate nonlinear state space framework, based on a discrete time
representation of the proposed model, is specified. Three measures constructed
from high frequency data, in addition to the daily return measure, are used to
define the multiple measurement equations. The high frequency measures
represent observed (price) jump occurrences and size, plus (logarithmic)
bipower variation. A Bayesian analysis of the model is undertaken using a
Markov chain Monte Carlo (MCMC) algorithm that accommodates the numerous
sources of non-linearity in the state space model, and\ that samples the
latent diffusion variances efficiently in blocks. The conditionally
deterministic (Hawkes) specification for the jump intensities is
computationally convenient, with the posterior distribution of both
intensities at any time point - including future time points - able to be
estimated from the MCMC draws of the parameters and latent variables to which
the intensities are functionally related.

Application of the methodology to data on the S\&P500\ index for the period
January 1996 to June 2014 is documented in detail. The empirical analysis
includes the calculation of marginal likelihoods for evaluating the proposed
specification against multiple alternatives, most of which are nested within
our general state space model and many of which share features with (or,
indeed, coincide with) models that have featured prominently in the
literature. Predictive distributions are also computed, for the purpose of
out-of-sample assessments. The comparative models include those in which the
price and variance jump intensities are dynamic as a consequence of a
functional dependence (either linear or non-linear) on the level of
volatility. Two realized generalized autoregressive conditional
heteroscedasticity (RGARCH) specifications of Hansen, Huang and Shek (2012)
are also entertained, as alternatives to the state space form.

As in Bandi and Reno (2016) spot price data only is used to analyse all
models, with the results unaffected as a consequence by the nature of - and
potential dynamics in - volatility and jump risk premia (see Bollerslev,
Gibson and Zhou, 2011, and Maneesoonthorn, Martin, Forbes and Grose, 2012, for
analyses in which such specifications do feature). However, and in contrast
with Bandi and Reno, data measured at the daily frequency (including that
which aggregates to the daily level over intraday observations) underpins the
analysis. In common with the large part of the relevant literature
(Bollerslev, Kretschmer, Pigorsch and\ Tauchen, 2009, and Liu, Patton and
Sheppard, 2015, amongst many others) we also choose to construct all measures
using within-day observations only, thereby avoiding the need to model
close-to-open movements in the index (as in,\textbf{ }for example, Ahoniemi,
Fuertes and Olmo, 2015, and Andersen, Bollerslev and Huang, 2011) and any
specific dynamic movements therein. (See Hansen and Lunde, 2005, and
Takahashi, Omori and Watanabe, 2009, for earlier discussions on the role
played by non-trading periods in the construction of high frequency measures).

The remainder of the paper is organized as follows. Section \ref{ch6svexj}
describes our proposed asset price model and its main properties. The
continuous time representation is presented first, followed by the discrete
time state space structure adopted for inference. Details are given of the
high frequency measures of volatility and price jumps that are used to
supplement daily returns in defining the state space model. The Bayesian
inferential approach is then outlined in Section \ref{ch6Sec:BayesInf},
including the way in which the alternative specifications are to be assessed,
relative to the most general model, both in terms of marginal likelihoods and
cumulative log scores. Results from the extensive empirical analysis of the
S\&P500 index are presented and discussed in Section \ref{ch6empirical}. The
benefits of allowing for a very flexible dynamic specification for price and
variance jumps are confirmed by both the within-sample and predictive
assessments, with the bivariate Hawkes specification given strong support by
the data, relative to other more restrictive models. The empirical results
also indicate that two jump intensity processes differ in terms of their time
series behaviour. Most notably, the variance jump intensity is much more
closely aligned with market conditions, exhibiting its most dramatic increase
at the peak of the global financial crisis in late 2008. Section
\ref{ch6concl7} provides some conclusions. Certain technical results,
including algorithmic and prior specification details, are included in
appendices to the paper.

\section{An asset price process with stochastic volatility and self-exciting
jumps \label{ch6svexj}}

\subsection{The continuous time representation\label{cont_time}}

Let $p_{t}=\ln\left(  P_{t}\right)  $ be the natural log\emph{\ }of the asset
price, $P_{t}$ at time\textbf{\ }$t>0$, whose evolution over time\textbf{\ }is
described by the following bivariate jump diffusion\textbf{\ }process,%
\begin{align}
dp_{t}  &  =\left(  \mu+\gamma V_{t}\right)  dt+\sqrt{V_{t}}dB_{t}^{p}%
+dJ_{t}^{p}\label{ch6pt}\\
dV_{t}  &  =\kappa\left(  \theta-V_{t}\right)  dt+\sigma_{v}\sqrt{V_{t}}%
dB_{t}^{v}+dJ_{t}^{v}, \label{ch6vt}%
\end{align}
with $B_{t}^{p}$ and $B_{t}^{v}$ denoting standard Brownian motion
processes,\textbf{\ }$corr(dB_{t}^{p},dB_{t}^{v})=\rho dt$ and $dJ_{t}%
^{i}=Z_{t}^{i}dN_{t}^{i}$, for $i=\left\{  p,v\right\}  $. Without the
discontinuous sample paths $dJ_{t}^{p}$ and $dJ_{t}^{v}$ this form of asset
pricing process replicates that of the Heston (1993) square root stochastic
volatility model, where the parameter restriction $\sigma_{v}^{2}\leq
2\kappa\theta$ ensures the positivity of the variance process, denoted by
$V_{t},$ for $t>0$. The drift component of (\ref{ch6pt}) contains the
additional component $\gamma V_{t}$, allowing for a volatility feedback effect
(that is, the impact of volatility on future returns) to be captured, while
$corr(dB_{t}^{p},dB_{t}^{v})=\rho dt$\ in (\ref{ch6vt}) captures the leverage
effect (that is, the impact of (negative) returns on future volatility). (See
Bollerslev, Livitnova and Tauchen, 2006, who also propose a model that
separates volatility feedback from leverage effects.) The\textbf{\ }$J_{t}%
^{i}$, $i=\left\{  p,v\right\}  ,$ are dependent\textbf{\ }random jump
processes that permit occasional jumps in either $p_{t}$ or $V_{t}$, or both,
and have random sizes $Z_{t}^{p}$ and $Z_{t}^{v}$, respectively.

A novel contribution of this paper is the specification of a bivariate Hawkes
process for the point processes, $N_{t}^{i},$\textbf{\ }$i=\left\{
p,v\right\}  $, which feeds into the bivariate jump process,\textbf{\ }%
$J_{t}^{i},$ $i=\left\{  p,v\right\}  .$\textbf{\ }Specifically, we assume
that\textrm{\ }%
\begin{align}
\Pr\left(  dN_{t}^{p}=1\right)   &  =\delta_{t}^{p}dt+o(dt)\text{,\ \ with
\ }\label{ch6_dnt}\\
d\delta_{t}^{p}  &  =\alpha_{p}\left(  \delta_{\infty}^{p}-\delta_{t}%
^{p}\right)  dt+\beta_{pp}dN_{t}^{p}, \label{ch6dlamp}%
\end{align}
and that\textbf{\ }%
\begin{align}
\Pr\left(  dN_{t}^{v}=1\right)   &  =\delta_{t}^{v}dt+o(dt)\text{,\ \ with
\ }\label{ch6dnt_2}\\
d\delta_{t}^{v}  &  =\alpha_{v}\left(  \delta_{\infty}^{v}-\delta_{t}%
^{v}\right)  dt+\beta_{vv}dN_{t}^{v}+\beta_{vp}dN_{t}^{p}+\beta_{vp}^{\left(
-\right)  }dN_{t}^{p\left(  -\right)  }, \label{ch6dlamv}%
\end{align}
where $dN_{t}^{p\left(  -\right)  }=dN_{t}^{p}\mathbf{1}\left(  Z_{t}%
^{p}<0\right)  $ denotes the\textbf{\ }occurrence of\textbf{\ }a
\textit{negative} price jump,\textbf{\ }corresponding to a value of one for
the indicator function\textbf{\ }$\mathbf{1}(\cdot).$ Due to the inclusion of
the terms $dN_{t}^{p}$ and $dN_{t}^{p\left(  -\right)  }$ in (\ref{ch6dlamv}),
the process $dN_{t}^{v}$ defined by (\ref{ch6dnt_2}) is not only
`self-exciting', but is also excited by a concurrent price jump. The
additional threshold component, $\beta_{vp}^{\left(  -\right)  }%
dN_{t}^{p\left(  -\right)  }$, allows a contemporaneous negative price
jump\textbf{\ }to have a differential impact (as compared with a positive
price jump)\textbf{ }on $d\delta_{t}^{v}$, thereby serving as an additional
channel for leverage, over and above the non-zero correlation between the
Brownian motion increments,\emph{\ }$dB_{t}^{p}$ and $dB_{t}^{v}.$ The
parameters $\delta_{\infty}^{i},$\ $i=\left\{  p,v\right\}  $, are the steady
state levels of the respective intensity processes to which the intensities
revert once the impact of excitation dissipates. See Hawkes (1971a,b) for
seminal discussions\textbf{\ }regarding self-exciting point processes, and
A\"{\i}t-Sahalia, Cacho-Diaz and Laeven\ (2015) for the introduction of the
Hawkes process into asset pricing models.

Our proposed specification can be viewed as a natural extension of the various
models in the literature that accommodate both stochastic volatility and
jumps. Most notably we relax the strict assumption of contemporaneous price
and volatility jumps as imposed, for example, by Duffie \textit{et al}.
(2000), Broadie \textit{et al}. (2007) and Bandi and Reno (2016). Instead,
price and volatility jumps are governed by separate, but dependent, dynamic
random processes, such that the two types of jumps may or may not coincide. As
detailed below, the probability of co-jumps can be readily computed from the
MCMC output, as can the posterior distributions for the magnitude of both
types of jumps (whether coincident or not). The specification can also be
viewed as an extension of the stochastic volatility model of A\"{\i}t-Sahalia
\textit{et al}. (2015), in which a Hawkes process is used to characterize
multivariate price jump occurrences, but with variance jumps absent from the
model. Similarly, it extends the model proposed by Fulop \textit{et al}.
(2014), in which price jump intensity (only) is characterized by a Hawkes
process, along with the restrictive assumption that variance jumps occur
contemporaneously with \textit{negative} price jumps.

\subsection{ A discrete time model for returns \label{Section:model}}

In common with the literature, we undertake inference in the context of a
discrete time state space representation of the continuous time model for the
asset price, applying an Euler discretization to (\ref{ch6pt}) through
(\ref{ch6dlamv}) with $\Delta t=1/252$ (equivalent to one\ trading day). Given
the complexity of the proposed model, and the multiple features on which we
wish to draw inference, we supplement the daily return measure, defined as
\[
r_{t}=p_{t+1}-p_{t},
\]
where $p_{t}$ denotes the logarithm of the asset price at the end of day $t$,
with three additional measures computed from high frequency\emph{
}(intraday)\textbf{ }returns. For expositional clarity we begin, in this
section, by focusing on the measurement equation for the return only,
describing in detail the latent components that feature therein. In Section
\ref{add_measures} we then introduce the high frequency quantities that are
used to define the three additional measurement equations, making clear the
assumed link between observed and latent quantities. In Section \ref{full_SSM}%
, we collect all components of the model together, introducing appropriate
labelling to facilitate subsequent referencing.

We begin then with the measurement equation based on the daily return,%
\begin{equation}
r_{t}=\mu+\gamma V_{t}+\sqrt{V_{t}}\xi_{t}^{p}+Z_{t}^{p}\Delta N_{t}^{p},
\label{rt_meas}%
\end{equation}
where the variation in\textbf{ }$r_{t}$ is driven by the latent diffusive
volatility process $V_{t}$ and the latent price jump component $Z_{t}%
^{p}\Delta N_{t}^{p}$. The (daily)\textbf{ }evolution of diffusive volatility
is given by%
\begin{equation}
V_{t+1}=\kappa\theta+\left(  1-\kappa\right)  V_{t}+\sigma_{v}\rho\left(
r_{t}-Z_{t}^{p}\Delta N_{t}^{p}-\mu-\gamma V_{t}\right)  +\sigma_{v}%
\sqrt{\left(  1-\rho^{2}\right)  V_{t}}\xi_{t}^{v}+Z_{t}^{v}\Delta N_{t}^{v},
\label{v_diff}%
\end{equation}
with the leverage parameter, $\rho$, taken into account explicitly.\textbf{\ }%
The error components $\xi_{t}^{p}$ and $\xi_{t}^{v}$, in (\ref{rt_meas}) and
(\ref{v_diff}),\ respectively, are\textbf{\ }defined as\textbf{\ }marginally
serially independent $N(0,1)$ sequences, with $corr\left(  \xi_{t}^{p},\xi
_{t}^{v}\right)  =0$\textbf{\ }for each\textbf{\ }$t$.

The latent occurrences of price and volatility jumps on day $t$ are expressed
as%
\begin{align}
\Delta N_{t}^{p}  &  \sim Bernoulli(\delta_{t}^{p})\label{delp}\\
\Delta N_{t}^{v}  &  \sim Bernoulli(\delta_{t}^{v}), \label{delv}%
\end{align}
with $\Delta N_{t}^{p}=N_{t+1}^{p}-N_{t}^{p}$, $\Delta N_{t}^{v}=N_{t+1}%
^{v}-N_{t}^{v},$ and where the probabilities of success are driven
(respectively) by the discretized intensity processes, \emph{ }%
\begin{align}
\delta_{t}^{p}  &  =\ \alpha_{p}\delta_{\infty}^{p}+\left(  1-\alpha
_{p}\right)  \delta_{t-1}^{p}+\beta_{pp}\Delta N_{t-1}^{p}\label{delta_p}\\
\delta_{t}^{v}  &  =\ \alpha_{v}\delta_{\infty}^{v}+\left(  1-\alpha
_{v}\right)  \delta_{t-1}^{v}+\beta_{vv}\Delta N_{t-1}^{v}+\beta_{vp}\Delta
N_{t-1}^{p}+\beta_{vp}^{\left(  -\right)  }\Delta N_{t-1}^{p\left(  -\right)
}. \label{delta_v}%
\end{align}
The discretized jump intensities, $\delta_{t}^{p}$ and $\delta_{t}^{v}$,
possess a conditionally deterministic structure that is analogous to that of a
generalized autoregressive conditional heteroskedastic (GARCH) model for
latent volatility, with the lagged jump occurrences playing a similar role to
the lagged (squared)\ returns in a GARCH model (Bollerslev, 1986). Assuming
stationarity, the\textbf{\ }unconditional mean for the price intensity process
is determined by taking expectations through (\ref{delta_p}) as follows,%
\[
E\left(  \delta_{t}^{p}\right)  =\ E\left(  \alpha_{p}\delta_{\infty}%
^{p}+\left(  1-\alpha_{p}\right)  \delta_{t-1}^{p}+\beta_{pp}\Delta
N_{t-1}^{p}\right)
\]
and solving for the common value $\delta_{0}^{p}=E\left(  \delta_{t}%
^{p}\right)  =E\left(  \delta_{t-1}^{p}\right)  =$ $E\left(  \Delta
N_{t-1}^{p}\right)  $ as%
\begin{equation}
\delta_{0}^{p}=\frac{\alpha_{p}\delta_{\infty}^{p}}{\alpha_{p}-\beta_{pp}}.
\label{delta_p_0}%
\end{equation}
Similarly, the unconditional mean of the variance jump intensity process in
(\ref{delta_v}) is given by\textbf{\ }$\delta_{0}^{v}=E\left(  \delta_{t}%
^{v}\right)  =E\left(  \delta_{t-1}^{v}\right)  =E\left(  \Delta N_{t-1}%
^{v}\right)  ,$ with%
\[
E\left(  \delta_{t}^{v}\right)  =\ E\left(  \alpha_{v}\delta_{\infty}%
^{v}+\left(  1-\alpha_{v}\right)  \delta_{t-1}^{v}+\beta_{vv}\Delta
N_{t-1}^{v}+\beta_{vp}\Delta N_{t-1}^{p}+\beta_{vp}^{\left(  -\right)  }\Delta
N_{t-1}^{p\left(  -\right)  }\right)  ,
\]
resulting in\textbf{\ }%
\begin{equation}
\delta_{0}^{v}=\frac{\alpha_{v}\delta_{\infty}^{v}+\beta_{vp}\delta_{0}%
^{p}+\beta_{vp}^{\left(  -\right)  }\pi_{p}\delta_{0}^{p}}{\alpha_{v}%
-\beta_{vv}}, \label{delta_v_0}%
\end{equation}
where $\pi_{p}=\Pr\left(  Z_{t}^{p}<0\right)  $ denotes the probability that
the price jump is negative. By substituting into equation (\ref{delta_v_0})
the expression for $\delta_{0}^{p}$ in (\ref{delta_p_0}), $\delta_{0}^{v}$ may
be re-expressed as the following function of static parameters,%
\[
\delta_{0}^{v}=\frac{\alpha_{v}\delta_{\infty}^{v}}{\alpha_{v}-\beta_{vv}%
}+\frac{\beta_{vp}\alpha_{p}\delta_{\infty}^{p}+\beta_{vp}^{\left(  -\right)
}\pi_{p}\alpha_{p}\delta_{\infty}^{p}}{\left(  \alpha_{v}-\beta_{vv}\right)
\left(  \alpha_{p}-\beta_{pp}\right)  }.
\]
To ensure that\textbf{ }$\delta_{0}^{p}\in\left(  0,1\right)  $\textbf{
}and\textbf{ }$\delta_{0}^{v}\in\left(  0,1\right)  $,\ the
restrictions\textbf{ }$0<\delta_{\infty}^{p}<\frac{\alpha_{p}-\beta_{pp}%
}{\alpha_{p}}$,\textbf{ }$0<\delta_{\infty}^{v}<\frac{\alpha_{v}-\beta
_{vv}-\beta_{vp}\delta_{0}^{p}-\beta_{vp}^{\left(  -\right)  }\pi_{p}%
\delta_{0}^{p}}{\alpha_{v}}$,\textbf{ }$0<\beta_{pp}<\alpha_{p}<1$\textbf{
}and\textbf{ }$0<\beta_{vv}<\alpha_{v}<1$\textbf{ }are required.\ Note that
the parameters used in (\ref{rt_meas})-(\ref{delta_v}) are the
\textit{discrete time} versions of the corresponding parameters in the
continuous time model (\ref{ch6pt})-(\ref{ch6dlamv}), but\textbf{ }with the
same symbols used for notational simplicity.

The size of the latent volatility jump is assumed to be exponentially
distributed,
\[
Z_{t}^{v}\sim Exponential\left(  \mu_{v}\right)  ,
\]
with only positive volatility jumps allowed as a consequence.\textbf{ }The
latent price jump size, on the other hand, is assumed to be composed of two
parts: magnitude and sign, ensuring adequate characterisation of the
empirically observed bimodal feature of the measured price jump distribution
(see further discussion of this point in Section \ref{ch6empirical}).
Specifically, we assume,%

\[
Z_{t}^{p}=S_{t}^{Z^{p}}\exp\left(  M_{t}^{p}\right)  ,
\]
where, with $\pi_{p}$ as defined earlier, the random variable,
\[
S_{t}^{Z^{p}}=\left\{
\begin{array}
[c]{cc}%
-1 & \text{ \ \ with probability }\pi_{p}\\
+1 & \text{ \ \ with probability }\left(  1-\pi_{p}\right)
\end{array}
\right.
\]
determines the sign of the price jump, and\textbf{ }%
\begin{equation}
M_{t}^{p}\sim N\left(  \mu_{p}+\gamma_{p}V_{t},\sigma_{p}^{2}\right)
\label{mt}%
\end{equation}
determines the logarithmic magnitude, with a mean value that is proportional
to the underlying volatility $V_{t}.$

The factors that drive the return in (\ref{rt_meas}) can be interpreted as
follows. Consistent with the empirical finance literature (see,\ for example,
Engle and\emph{\ }Ng, 1993, Maheu\ and McCurdy, 2004, and Malik, 2011), the
diffusive price shock, $\sqrt{V_{t}}\xi_{t}^{p},$ and the price jump
occurrence, $Z_{t}^{p}\Delta N_{t}^{p},$ are collectively viewed as `news'.
Regular modest movements in price, as driven by $\sqrt{V_{t}}\xi_{t}^{p}$, are
assumed to result from typical daily information flows, with (all else
equal)\textbf{\ }the typical direction\textbf{\ }of the impact of $\sqrt
{V_{t}}\xi_{t}^{p}$ on the variance of the subsequent period, $V_{t+1}$,
captured by the sign of $\rho$. The occurrence of a price jump however,
indicated by $\Delta N_{t}^{p}=1$, can be viewed as a sizably\emph{\ }%
larger\ than expected shock\textsf{\ }that may signal a shift in market
conditions, with the probability of subsequent price and/or variance jumps
adjusted accordingly, through the model adopted here for the jump intensities.
That is, the process $\Delta N_{t}^{p}$ can be viewed as being potentially
self-exciting: provoking an increase in the future intensity (and thus
occurrence) of price jumps (via (\ref{delta_p})), as well as provoking (or
exciting) an increase in the future intensity of variance jumps (via
(\ref{delta_v})). The threshold parameter $\beta_{vp}^{\left(  -\right)  }%
$\ in (\ref{delta_v}) allows for a possible additional impact of a negative
price jump on the variance jump intensity (and, hence, the level of
volatility), providing an additional channel for leverage, as noted
above.\textbf{\ }

From the form of (\ref{rt_meas}) and (\ref{v_diff}), the implications for
returns of the occurrence of the two types of jumps are clear. From
(\ref{rt_meas}), the direct impact of a given price jump at time $t$, $\Delta
J_{t}^{p}=Z_{t}^{p}\Delta N_{t}^{p}$, is felt only at time $t$. However,
clusters of price jumps and, hence, successive extreme values in returns, can
occur via the dynamic intensity process in (\ref{delta_p}) that drives
subsequent realizations of $\Delta N_{t}^{p}$. The impact of a given
(positive) variance jump at time $t$ will carry forward through time via the
persistence of the $V_{t+1}$ process, as governed by $\kappa$. That is, if the
return\textbf{\ }variance jumps in any period, it will tend to\ remain higher
in subsequent periods and, thus, be expected to cause larger movements in
successive prices than would otherwise have occurred. In addition, any
clustering of variance jumps, driven by the dynamic intensity process in
(\ref{delta_v}), would simply cause an exaggeration of the resultant
clustering of extreme returns.\textbf{\ }Arguably, clusters of jumps in the
latent variance would typically be associated with sustained market
instability, with the variance jump intensity expected to increase and remain
high during periods of heightened market stress. We return to this point in
Section \ref{ch6empirical}.

\subsection{Incorporating high frequency measurements of volatility and price
jumps \label{add_measures}}

In the spirit of Barndorff-Nielsen and\ Shephard (2002), Creal (2008),
Takahashi\textbf{ }\textit{et al.\textbf{ }}(2009), Dobrev and Szerszen
(2010), Jacquier and\ Miller (2010), Hansen \textit{et al.} (2012),
Maneesoonthorn \textit{et al.} (2012), and Koopman and Scharth (2013), amongst
others, we exploit high-frequency data to supplement the measurement equation
in (\ref{rt_meas}) with additional equations\ based on nonparametric measures
of return variation: both its diffusive and jump components. As is now
standard knowledge, realized variance, defined by \emph{\ }%
\begin{equation}
RV_{t}=%
{\textstyle\sum\limits_{t<t_{i}\leq{t+1}}^{M}}
r_{t_{i}}^{2}, \label{ch6RV}%
\end{equation}
where\emph{\ }$r_{t_{i}}=p_{t_{i+1}}-p_{t_{i}}$ denotes the $i^{th}$ observed
return the over the horizon $t$ to $t+1$, and there being $M$ such returns, is
a consistent estimator of quadratic variation under the assumption of no
microstructure noise. (See, for example, Barndorff-Nielsen and Shephard, 2002
and Andersen, Bollerslev, Diebold and Labys,\ 2003). Quadratic variation,
$Q\mathcal{V}_{t,t+1}$, in turn,\textbf{ }captures the variation in the return
over the horizon $t$ to $t+1$ due to both the stochastic volatility and price
jump components, with $Q\mathcal{V}_{t,t+1}=\mathcal{V}_{t,t+1}+\mathcal{J}%
_{t,t+1}^{2},$ where $\mathcal{V}_{t,t+1}=\int_{t}^{t+1}V_{s}ds$ denotes the
integrated variance, and$\mathcal{\ J}_{t,t+1}^{2}=\sum_{t<s\leq t+1}%
^{N_{t+1}^{p}}\left(  Z_{s}^{p}\right)  ^{2}$ denotes the price jump
variation. With bipower variation,
\begin{equation}
BV_{t}=\frac{\pi}{2}%
{\textstyle\sum\limits_{t<t_{i}\leq{t+1}}^{M}}
\left\vert r_{t_{i}}\right\vert \left\vert r_{t_{i-1}}\right\vert ,
\label{chp6BV}%
\end{equation}
being a consistent measure of $\mathcal{V}_{t,t+1}$ (again, in the absence of
microstructure noise), the discrepancy between $RV_{t}$ in (\ref{ch6RV}) and
$BV_{t}$ in (\ref{chp6BV}) serves as a measure of price jump variation, and
has, as a consequence, formed the basis of various tests of the significance
of jump variation on any particular day; see, for example, Barnorff-Nielsen
and Shephard (2004, 2006) and Huang and Tauchen (2005).

Three measurement equations that exploit the information content in $RV_{t}$
and $BV_{t}$ are constructed as follows. First we define a measure that
indicates the \textit{occurrence} or otherwise of a price jump on day $t$,%
\begin{equation}
I_{t}^{p}=\mathbf{1}\left(  Z_{RJ,t}>c_{a}\right)  , \label{ipt}%
\end{equation}
where%
\begin{equation}
Z_{RJ,t}=\frac{RJ_{t}}{\sqrt{\left(  \frac{\pi^{2}}{4}+\pi-5\right)
M^{-1}\max\left(  1,\frac{TQ_{t}}{BV_{t}^{2}}\right)  }}, \label{zrj}%
\end{equation}
$RJ_{t}=\left(  RV_{t}-BV_{t}\right)  /RV_{t}$ and $c_{a}=\Phi^{-1}\left(
1-a\right)  $ is the critical value in\textbf{ }a standard normal
distribution, associated with significance level $a.$ The term $TQ_{t}$ in the
denominator of (\ref{zrj}) denotes an estimate of the integrated quarticity,
with $Z_{RJ,t}$ having\textbf{ }a limiting standard normal distribution under
the assumption of no jumps, as a result of the particular standardization used
in its definition; see Huang and Tauchen (2005)\textbf{ }for details.\textbf{
}The indicator function in (\ref{ipt}) is then viewed as a noisy measure of
the latent price jump occurrence in (\ref{delp}). That is, we specify the
measurement equation,%
\[
I_{t}^{p}=\left\{
\begin{array}
[c]{cc}%
Bernoulli\left(  \beta\right)  & \text{ \ \ if }\Delta N_{t}^{p}=1\\
Bernoulli\left(  \alpha\right)  & \text{ \ \ if }\Delta N_{t}^{p}=0
\end{array}
\right.  .
\]
with constant probabilities $\alpha$ and $\beta$ to be estimated from the data.

Second, we assume that the latent (logarithmic) price jump size in (\ref{mt})
is measured with error by%
\begin{equation}
\widetilde{M}_{t}^{p}=\ln\left(  \widetilde{Z}_{t}^{p}\right)  , \label{Mpt}%
\end{equation}
where%
\begin{equation}
\widetilde{Z}_{t}^{p}=\sqrt{\max\left(  RV_{t}-BV_{t},0\right)  }, \label{zp}%
\end{equation}
by specifying the measurement equation,
\[
\widetilde{M}_{t}^{p}=M_{t}^{p}+\sigma_{M_{p}}\xi_{t}^{M_{p}}%
\ \ \ \ \ \text{for }\widetilde{Z}_{t}^{p}\neq0,
\]
with\textbf{ }$\xi_{t}^{M_{p}}\sim i.i.d.N\left(  0,1\right)  .$\footnote{Note
that when $\widetilde{Z}_{t}^{p}=0$, which, from (\ref{zp}), occurs when
$RV_{t}-BV_{t}\leq0$, we do not view the data as providing any information
about price jump size, and with $\widetilde{M}_{t}^{p}$ being undefined in
this case.}

Third, as a direct measure of integrated volatility, $BV_{t}$ is assumed to
bring noisy information about the diffusive volatility process, including any
jumps in such a process. Hence (and utilizing $BV_{t}$ in logarithmic form in
order to better justify the assumption of a Gaussian measurement error), we
specify a final measurement equation as\textbf{ }
\[
\ln BV_{t}=\psi_{0}+\psi_{1}\ln V_{t}+\sigma_{BV}\xi_{t}^{BV},
\]
where $\xi_{t}^{BV}\sim i.i.d.N\left(  0,1\right)  $, where $\ln V_{t}$ is a
discretization of\textbf{ }$\ln\mathcal{V}_{t,t+1}$\textbf{ }and the
estimation of\textbf{ }$\psi_{0}$ and $\psi_{1}$ as free parameters allows for
$\ln BV_{t}$ to be a biased measure of\textbf{ }$\ln V_{t}.$

\subsection{The full discrete time state space model\label{full_SSM}}

For expositional clarity, we collect together here all components of the
model, beginning with the four measurement equations:
\begin{align}
\text{\textit{Daily return}}  &  \text{:\textbf{\quad}}r_{t}=\mu+\gamma
V_{t}+\sqrt{V_{t}}\xi_{t}^{p}+Z_{t}^{p}\Delta N_{t}^{p}\label{m1}\\
\text{\textit{Price jump indicator}}  &  \text{:\textbf{\quad}}I_{t}%
^{p}=\left\{
\begin{array}
[c]{cc}%
Bernoulli\left(  \beta\right)  & \text{ \ \ if }\Delta N_{t}^{p}=1\\
Bernoulli\left(  \alpha\right)  & \text{ \ \ if }\Delta N_{t}^{p}=0
\end{array}
\right. \label{m2}\\
\text{\textit{Log price jump size}}  &  \text{:\textbf{\quad}}\widetilde{M}%
_{t}^{p}=M_{t}^{p}+\sigma_{M_{p}}\xi_{t}^{M_{p}}\ \ \ \ \ \text{(for
}\widetilde{Z}_{t}^{p}\neq0\text{)}\label{m3}\\
\text{\textit{Log bipower\textbf{ }variation}}  &  \text{:\textbf{\quad}}\ln
BV_{t}=\psi_{0}+\psi_{1}\ln V_{t}+\sigma_{BV}\xi_{t}^{BV}. \label{m4}%
\end{align}
The stochastic state processes comprise:%
\begin{align}
\text{\textit{Latent volatility}}  &  \text{:\textbf{\quad}}V_{t+1}%
=\kappa\theta+\left(  1-\kappa\right)  V_{t}+\sigma_{v}\rho\left(  r_{t}%
-Z_{t}^{p}\Delta N_{t}^{p}-\mu-\gamma V_{t}\right) \nonumber\\
&  \text{\hspace{0.6in}}+\sigma_{v}\sqrt{\left(  1-\rho^{2}\right)  V_{t}}%
\xi_{t}^{v}+Z_{t}^{v}\Delta N_{t}^{v}\label{s1}\\
\text{\textit{Latent price jump occurrence}}  &  \text{:\textbf{\quad}}\Delta
N_{t}^{p}\sim Bernoulli(\delta_{t}^{p})\label{s2}\\
\text{\textit{Latent volatility jump occurence}}  &  \text{:\textbf{\quad}%
}\Delta N_{t}^{v}\sim Bernoulli(\delta_{t}^{v})\label{s3}\\
\text{\textit{Latent price jump size}}  &  \text{:\quad}Z_{t}^{p}=S_{t}%
^{Z^{p}}\exp\left(  M_{t}^{p}\right) \label{s4}\\
\text{\textit{Latent volatility jump size}}  &  \text{:\quad}Z_{t}^{v}\sim
Exponential\left(  \mu_{v}\right)  , \label{s5}%
\end{align}
where the specification of the latent price jump $Z_{t}^{p}$ in (\ref{s4}) is
given by the product of two random components, with one relating to the jump
direction:%
\begin{equation}
S_{t}^{Z^{p}}=\left\{
\begin{array}
[c]{cc}%
-1 & \text{ \ \ with probability }\pi_{p}\\
+1 & \text{ \ \ with probability }\left(  1-\pi_{p}\right)
\end{array}
\right. \nonumber
\end{equation}
and the other relating to the log price jump magnitude:\textbf{ }%
\begin{equation}
M_{t}^{p}\sim N\left(  \mu_{p}+\gamma_{p}V_{t},\sigma_{p}^{2}\right)
.\nonumber
\end{equation}
Finally, the two conditionally deterministic states are given by:%
\begin{align}
\text{\textit{Price jump intensity}}  &  \text{:\quad}\delta_{t}^{p}%
=\alpha_{p}\delta_{\infty}^{p}+\left(  1-\alpha_{p}\right)  \delta_{t-1}%
^{p}+\beta_{pp}\Delta N_{t-1}^{p}\label{d1}\\
\text{\textit{Volatility jump intensity}}  &  \text{:\quad}\delta_{t}%
^{v}=\alpha_{v}\delta_{\infty}^{v}+\left(  1-\alpha_{v}\right)  \delta
_{t-1}^{v}+\beta_{vv}\Delta N_{t-1}^{v}\nonumber\\
&  \text{\hspace{0.6in}}+\beta_{vp}\Delta N_{t-1}^{p}+\beta_{vp}^{\left(
-\right)  }\Delta N_{t-1}^{p\left(  -\right)  }. \label{d2}%
\end{align}
All subsequent referencing of the model makes use of the equation numbering in
this section.

\section{Bayesian inference\label{ch6Sec:BayesInf}}

\subsection{Overview}

For notational convenience, we collectively denote, at time point $t$, the
measurement vector as $Y_{t}=\left(  r_{t},I_{t}^{p},\widetilde{M}_{t}^{p},\ln
BV_{t}\right)  ^{\prime},$ and the latent state vector as $X_{t}=\left(
V_{t},\Delta N_{t}^{p},\Delta N_{t}^{v},S_{t}^{Z_{p}},M_{t}^{p},Z_{t}%
^{v}\right)  ^{\prime}$, with the static parameters also collectively denoted
by the vector
\begin{equation}
\phi=(\mu,\gamma,\rho,\mu_{p},\gamma_{p},\sigma_{p},\pi_{p},\alpha
,\beta,\sigma_{M_{p}},\psi_{0},\psi_{1},\sigma_{BV},\kappa,\theta,\sigma
_{v},\mu_{v},\delta_{0}^{p},\alpha_{p},\beta_{pp},\delta_{0}^{v},\alpha
_{v},\beta_{vv},\beta_{vp},\beta_{vp}^{\left(  -\right)  })^{\prime}.
\label{phi}%
\end{equation}
In addition, we denote time-indexed\textbf{\ }variables generically as, for
example, $W_{1:t}=\left(  W_{1},...,W_{t}\right)  ^{\prime}$ for $t=1,...,T$,
where $W_{1:0}$ is empty. The joint posterior density associated with the full
model in (\ref{m1})-(\ref{d2}), denoted subsequently by $\mathcal{M}_{F}$,
satisfies%
\begin{equation}
p\left(  X_{1:T},\phi\mathbf{|}Y_{1:T}\right)  \propto p\left(  Y_{1}%
|X_{1},\phi\right)  p\left(  X_{1}|\phi\right)  p\left(  \phi\right)  \left[
{\textstyle\prod\limits_{t=2}^{T}}
p\left(  Y_{t}|X_{1:t-1,}\phi\right)  \times p\left(  X_{t}|X_{1:t-1,}%
\phi\right)  \right]  . \label{post}%
\end{equation}
Note that this joint posterior assumes that $\delta_{1}^{p}=\delta_{0}^{p},$
$\delta_{1}^{v}=\delta_{0}^{v}$ and $\Delta N_{1}^{v}=\Delta N_{1}^{p}$. In
the specification of the prior $p\left(  \phi\right)  $ in (\ref{post}) we use
a combination of noninformative and weakly informative distributions for the
various elements of $\phi$. Other than exploiting the natural groupings of
parameters that arise from the regression structures embedded within the
model, we adopt \textit{a priori} independence for the individual parameters.
The detailed\textbf{ }specifications of each component of\textbf{ }$p\left(
\phi\right)  $ are documented in Appendix A.

Given the complexity of the state space representation, and the high dimension
of the set of unknowns, the posterior indicated by\textbf{ }(\ref{post}%
)\textbf{ }is not available in closed form. Hence, a hybrid of the Gibbs and
Metropolis-Hastings (MH) MCMC algorithms is developed to obtain draws of the
static parameters and latent variables of interest from the joint posterior
distribution, with inference - including the construction of posterior
predictive distributions - conducted using those draws. Details of this
algorithm, including a reference made to Maneesoonthorn\ \textit{et al.}
(2012) for a full\textbf{ }description of the multi-move algorithm adopted for
sampling the variance state vector, $V_{1:T}$, are given in Appendix B.1.

\subsection{Models of interest and their marginal likelihoods\label{bayes}}

As has been highlighted, a novel aspect of our specification is that it allows
for dynamic behaviour in both price and variance jumps, as well as various
types of dependencies between those extreme movements.\ It is of interest then
to explore whether or not this rich dynamic structure is warranted
empirically, through an investigation of various simpler specifications.

To this end, we consider several competing models, summarized in Table
\ref{models}%
, most of which are nested in the full model\textbf{ }specification
$\mathcal{M}_{F}$ in (\ref{m1}) to (\ref{d2}), and all of which are to be
evaluated empirically in Section \ref{ch6empirical}. First (and with reference
to the labelling of models in the left-most column of the table), a model
without\textbf{\emph{ }}a threshold component (that is, without the
differential impact on variance jump intensity due to\emph{ }the occurrence of
\textit{negative }price jumps) is specified as $\mathcal{M}_{1}$. Next, model
$\mathcal{M}_{2}$ specifies that the occurrence of price jumps has no impact
at all on the variance jump intensity, via the removal of both price jump
feedback terms from $\delta_{t}^{v}$. That price and variance jumps occur
contemporaneously, or\textbf{ }that the variance does not jump at all, each
correspond to further\textbf{ }restrictions specified in $\mathcal{M}_{3}$ and
$\mathcal{M}_{4}$, respectively. Note that $\mathcal{M}_{4}$ shares some
features with the model proposed by A\"{\i}t-Sahalia \textit{et al.} (2015),
albeit in a single asset setting here. In models $\mathcal{M}_{5}$ to
$\mathcal{M}_{7}$, and as an alternative to the use of the bivariate Hawkes
process, the jump intensities are specified as various functions (both linear
and non-linear) of the latent variance, with $\mathcal{M}_{5}$ sharing some
common features with the models adopted in Bates (1996), Pan (2002), and
Eraker (2004). In $\mathcal{M}_{8}$ we then specify constant jump intensities,
yielding the stochastic volatility with the independent jumps (SVIJ) model of
Duffie \textit{et al.} (2000), whilst in $\mathcal{M}_{9}$ we\ consider the
absence of both price and variance jumps, with the resultant latent process
thereby coinciding with that of the conventional Heston (1993) square root
model.\emph{ }

Finally, we provide an alternative to the state space form, entertaining the
conditionally deterministic RGARCH(1,1) model of Hansen\textit{ et al}.
(2012), specified as%
\begin{align}
r_{t}  &  =\sqrt{h_{t}}z_{t}\nonumber\\
h_{t}  &  =v\left(  h_{t-1},BV_{t-1}\right) \label{RGARCH2}\\
BV_{t}  &  =m\left(  h_{t},z_{t},u_{t}\right)  , \label{RGARCH3}%
\end{align}
where $z_{t}\sim N(0,1)$ and $u_{t}\sim N(0,\sigma_{u}^{2})$, with $v\left(
.\right)  $ and $m\left(  .\right)  $ defining the evolution of the
deterministic variance and the bipower variation, respectively. In our
comparison, we entertain both the linear and the log-linear specifications of
RGARCH, denoted by $\mathcal{M}_{10}$ and $\mathcal{M}_{11},$ respectively,
with details given in Table
\ref{models}%
. These two models are, of course, not nested in the general state space
framework, and details of the separate MCMC algorithm used to estimate the
relevant posterior densities are provided in Appendix B.2.

To examine\emph{ }the relative merits of\textbf{ }these twelve models of
interest, computation of their corresponding marginal likelihood values,
\begin{equation}
p\left(  Y_{1:T}|\mathcal{M}_{i}\right)  , \label{ch6marg_lik}%
\end{equation}
for $i=F$ and $i=1,...,11,$ is required. Under the assumption that each of the
models is \textit{a priori }equally likely, the posterior odds ratio for the
full state space model $\mathcal{M}_{F}$ relative to\textbf{ }any restricted
model $\mathcal{M}_{i}$ is equivalent to the Bayes factor $BF_{i}$, given in
turn by
\begin{equation}
BF_{i}=\frac{p\left(  Y_{1:T}|\mathcal{M}_{F}\right)  }{p\left(
Y_{1:T}|\mathcal{M}_{i}\right)  }. \label{ch6bf_if}%
\end{equation}
Given Bayes factors $BF_{i}$ and $BF_{j}$, the Bayes factor for model
$\mathcal{M}_{i}$ relative to model $\mathcal{M}_{j}$ is obtained simply as
$BF_{i,j}=BF_{j}/BF_{i}.$\ 

Note that both $\mathcal{M}_{F}$ and the first eight comparator models are
estimated using the full set of measurements. However, models $\mathcal{M}%
_{9},$ $\mathcal{M}_{10}$ and $\mathcal{M}_{11}$ do not exploit observed price
jump information, given the absence of any jumps specified in either the price
or latent volatility component, with all three models estimated using only
observations on $r_{t}$ and $BV_{t}$ as a consequence. This gives us two
possibilities regarding the computation of the Bayes factors for these models.
Firstly, we can compute the marginal likelihoods using only observations on
$r_{t}$ and $BV_{t}$ and compare the three marginal likelihoods to each other
only. Alternatively, we can augment these marginal likelihoods with an
additional factor that caters for the price jump measurements, computed using
priors that are consistent with the imposition of no jumps within the models.
These latter (expanded) quantities can then be used in a comparison, via the
computation of Bayes factors, with the other eight specifications. We record
both forms of results in Section \ref{ch6empirical}.%

\begin{table}[tbhp] \centering
\caption
{Specification of the full set of models used in the comparative evaluation. All parametric restrictions described herein relate to the parameters of
either equations (38) and (39) or equations (43) and (44).}
\label{models}%
%

\begin{tabular}
[c]{lll}
&  & \\
Model & Restrictions & Description\\\hline\hline
&  & \\
$\mathcal{M}_{F}$ & None$\smallskip$ & Full state space model: (\ref{m1}) to
(\ref{d2})\\
$\mathcal{M}_{1}$ & $\beta_{vp}^{\left(  -\right)  }=0\smallskip$ & Full model
without threshold component\\
$\mathcal{M}_{2}$ & $\beta_{vp}^{\left(  -\right)  }=\beta_{vp}=0\smallskip$ &
Full model without price jump feedback\\
$\mathcal{M}_{3}$ & $\Delta N_{t}^{p}=\Delta N_{t}^{v}\smallskip$ & Full model
with contemporaneous jumps\\
$\mathcal{M}_{4}$ & $\Delta N_{t}^{v}=0\smallskip$ & Hawkes model without
volatility jumps\\
$\mathcal{M}_{5}$ & $\left\{
\begin{tabular}
[c]{l}%
$\delta_{t}^{p}=\alpha_{p_{0}}+\alpha_{p_{1}}V_{t}$\\
$\delta_{t}^{v}=\alpha_{v_{0}}+\alpha_{v_{1}}V_{t}$%
\end{tabular}
\right.  \smallskip$ & State dependent jump intensity: linear\\
$\mathcal{M}_{6}$ & $\left\{
\begin{tabular}
[c]{l}%
$\delta_{t}^{p}=\alpha_{p_{0}}+\alpha_{p_{1}}V_{t}+\alpha_{p_{2}}V_{t}^{2}$\\
$\delta_{t}^{p}=\alpha_{v_{0}}+\alpha_{v_{1}}V_{t}+\alpha_{v_{2}}V_{t}^{2}$%
\end{tabular}
\right.  $ & State dependent jump intensity: quadratic\\
$\mathcal{M}_{7}$ & $\left\{
\begin{tabular}
[c]{l}%
$\delta_{t}^{p}=\frac{\exp\left(  \alpha_{p_{0}}+\alpha_{p_{1}}V_{t}\right)
}{1+\exp\left(  \alpha_{p_{0}}+\alpha_{p_{1}}V_{t}\right)  }$\\
$\delta_{t}^{v}=\frac{\exp\left(  \alpha_{v_{0}}+\alpha_{v_{1}}V_{t}\right)
}{1+\exp\left(  \alpha_{v_{0}}+\alpha_{v_{1}}V_{t}\right)  }$%
\end{tabular}
\right.  $ & State dependent jump intensity: logistic\\
$\mathcal{M}_{8}$ & $\delta_{t}^{p}=\delta_{0}^{p}\text{, }\delta_{t}%
^{v}=\delta_{0}^{v}\smallskip$ & Constant\textbf{ }jump intensity\\
$\mathcal{M}_{9}$ & $\delta_{t}^{p}=0\text{ and }\delta_{t}^{v}=0\smallskip$ &
Stochastic volatility model without jumps\\
$\mathcal{M}_{10}$ & $\left\{
\begin{tabular}
[c]{l}%
$h_{t}=\omega+\beta h_{t-1}+\gamma BV_{t-1}$\\
$BV_{t}=\xi+\varphi h_{t}+\tau_{1}z_{t}$\\
$\qquad+\tau_{2}\left(  z_{t}^{2}-1\right)  +u_{t}$%
\end{tabular}
\right.  $ & RGARCH model: linear\\
$\mathcal{M}_{11}$ & $\left\{
\begin{tabular}
[c]{l}%
$\ln h_{t}=\omega+\beta\ln h_{t-1}+\gamma\ln BV_{t-1}$\\
$\ln BV_{t}=\xi+\varphi\ln h_{t}+\tau_{1}z_{t}$\\
$\qquad+\tau_{2}\left(  z_{t}^{2}-1\right)  +u_{t}$%
\end{tabular}
\right.  $ & RGARCH model: log-linear\\
&  & \\\hline\hline
\end{tabular}
%

\end{table}%

The marginal likelihood for model $\mathcal{M}_{i}$\ in (\ref{ch6marg_lik}) is
challenging to compute, in particular for the state space specifications,
which require\textbf{ }the calculation of an integral over a very large
dimension due to the number of latent variables present. As per Chib (1995)
and\ Chib and\ Jeliazkov (2001), we estimate the marginal likelihood of each
model using the output of a series of auxiliary\textbf{ }MCMC algorithms, in
addition to the full MCMC algorithm associated with estimation of the given
model. Specific details of this computation are provided in Appendix C. A
brief explanation of the computation of the marginal likelihoods for the three
restricted models ($\mathcal{M}_{9},$ $\mathcal{M}_{10}$ and $\mathcal{M}%
_{11}$) is also provided in this appendix.

\subsection{Predictive performance\label{pred_perf}}

With reference to the \textit{joint }measurement vector at time $t$, and model
$\mathcal{M}_{i}$, $i=F$ and $i=1,...,11$, the one-step-ahead predictive
distribution as based on information up to time $t-1$ is given by%
\begin{align}
&  p\left(  Y_{t}|Y_{1:t-1},\mathcal{M}_{i}\right) \label{pred}\\
&  =\int p\left(  Y_{t}|Y_{1:t-1},X_{1:t},\phi_{i},\mathcal{M}_{i}\right)
p\left(  X_{1:t}|Y_{1:t-1},\phi_{i},\mathcal{M}_{i}\right)  p\left(  \phi
_{i}|Y_{1:t-1},\mathcal{M}_{i}\right)  dX_{1:t}d\phi_{i},\nonumber
\end{align}
where $\phi_{i}$ denotes the vector of static parameters associated with model
$\mathcal{M}_{i}$, and $X_{1:t}$ represents the full set of latent variables
that feature therein. For the RGARCH models, of course, $X_{1:t}$ is an empty
set, so that the integration occurs over $\phi_{i}$ only. As is well known
(see, for example, Geweke, 2001) the log marginal likelihood for any model
$\mathcal{M}_{i}$, $i=F$ and $i=1,...,11$, computed over the entire sample
period 1 to $T$,\textbf{ }may be expressed as the sum of $T$ log marginal
\textit{predictive} densities, each associated with $\mathcal{M}_{i}$, and
evaluated\ \textit{ex-post} at the corresponding observed values:%
\begin{equation}
\ln p\left(  Y_{1:T}|\mathcal{M}_{i}\right)  =\sum_{t=1}^{T}\ln p\left(
Y_{t}|Y_{1:t-1},\mathcal{M}_{i}\right)  . \label{logmarg_decomp}%
\end{equation}
It follows that the Bayes factor $BF_{i}$ in (\ref{ch6bf_if}) may actually be
interpreted as providing a measure of predictive accuracy for model
$\mathcal{M}_{F}$, relative to that of model $\mathcal{M}_{i}$, over the
entire sample period. Importantly, the component predictive distributions in
(\ref{logmarg_decomp}) do not rely upon any unknown parameters, and reflect
the evaluation of predictions made without reference to any future information.

What is absent from the computation of the full sample Bayes factor, however,
is any information on the\textbf{ }\textit{change }over the sample period in
the predictive performance $\mathcal{M}_{F}$ relative to $\mathcal{M}_{i}$. To
better capture this dynamic predictive behaviour, we compute the cumulative
difference in log score (CLS) over an evaluation (sub-) period of $\left(
T-T_{0}\right)  $ trading days, according to
\begin{equation}
CLS_{i}\left(  n\right)  =\sum\limits_{t=T_{0}+1}^{n}\ln\left[  \frac{p\left(
Y_{t}|Y_{1:t-1},\mathcal{M}_{F}\right)  }{p\left(  Y_{t}|Y_{1:t-1}%
,\mathcal{M}_{i}\right)  }\right]  , \label{cls_full}%
\end{equation}
for $n=T_{0}+1,...,T$. An increase in the value of $CLS_{i}(n)$, relative to
$CLS_{i}(n-1),$ indicates an improvement in the performance of the reference
model, $\mathcal{M}_{F},$\textbf{ }relative to\textbf{ }$\mathcal{M}_{i}$, in
terms of predicting all four elements of $Y_{n}$, with a \textit{persistent}
positive level in $CLS_{i}$ indicating sustained predictive superiority of
$\mathcal{M}_{F}$ relative to $\mathcal{M}_{i}$, over that period. Note that
for the early values in this set of sequential $CLS_{i}(n)$ calculations to be
reliable, an\textbf{ }initial sample consisting of\textbf{ }$T_{0}$
observations, $Y_{1:T_{0}}$, is used to initialise the computation. (See also
Geweke and\ Amisano, 2010). Estimation of (\ref{pred}) and, subsequently,
computation of (\ref{cls_full}), occurs via a combination of MCMC and particle
filtering algorithms, with further details provided in Section \ref{pred_eval}.

To complement the CLS results as they pertain to the joint measurement vector
$Y_{t}$, in Section \ref{pred_eval} we also report predictive performance as
it relates to certain individual sub-vectors of $Y_{t}$. Denote any sub-vector
of $Y_{t}$ by $g_{t}$. Then we define the \textit{marginal CLS} for $g_{t}$
as\textbf{ }%
\begin{equation}
\lbrack g]\text{ }CLS_{i}\left(  n\right)  =\sum\limits_{t=T_{0}+1}^{n}%
\ln\left[  \frac{p\left(  g_{t}|Y_{1:t-1},\mathcal{M}_{F}\right)  }{p\left(
g_{t}|Y_{1:t-1},\mathcal{M}_{i}\right)  }\right]  , \label{marg_cls}%
\end{equation}
again for $n=T_{0}+1,...,T$. Specifically, we report predictive results for
the return, where $g_{t}=r_{t},$ and the log bipower variation, where
$g_{t}=\ln BV_{t}$, as well as for the\textbf{ }bi-variate subvector
containing the price\emph{ }jump indicator and size components, where%
\begin{equation}
g_{t}=\left(  \widetilde{M}_{t}^{p},I_{t}^{p}\right)  ^{\prime}.
\label{g_predpricejump}%
\end{equation}
Computation of the terms comprising the marginal CLS in (\ref{marg_cls}) for a
given $g_{t}$ requires a relatively small modification of the methodology used
to compute the joint quantities in (\ref{pred}) and (\ref{cls_full}), with
details provided in Section \ref{pred_eval}. Isolation of the predictives for
individual elements of $Y_{t}$ also enables us to directly compare the
predictive performance of models $\mathcal{M}_{9},$ $\mathcal{M}_{10}$ and
$\mathcal{M}_{11}$ with $\mathcal{M}_{F}$, in terms of the accuracy with which
these two models forecast future values of $g_{t}=r_{t}$ and $g_{t}=\ln
BV_{t}$ specifically. In the case of $\mathcal{M}_{10}$ and $\mathcal{M}_{11}%
$, given the absence of stochastic latent variables, computation of the
predictive quantities using the MCMC draws from the joint posterior is
standard, without there being any need for additional filtering steps.

\section{Empirical application \label{ch6empirical}}

\subsection{Data description and preliminary analysis}

For the empirical analysis documented below, 4598 observations on the
open-to-close logarithmic S\&P500 return ($r_{t}$), price jump indicator
($I_{t}^{p}$), logarithmic\textbf{ }price jump size $\left(  \widetilde{M}%
_{t}^{p}\right)  $ and logarithmic bipower variation ($\ln BV_{t}$) were
analyzed, over the period January 3, 1996 to June 23, 2014. The index data has
been supplied by the Securities Industries Research Centre of Asia Pacific
(SIRCA) on behalf of Reuters, with the raw intraday index data having been
cleaned using methods similar to those of Brownlees and Gallo (2006). The
measures constructed from high-frequency data are based on fixed five minute
sampling, with a `nearest price' method (Andersen, Bollerslev and Diebold,
2007) applied to construct the relevant returns five minutes apart, and only
index values recorded within the New York Stock Exchange market trading hours.
The numerical results reported in this empirical section have been produced
using a combination of the JAVA and MATLAB programming languages. Marginal
posterior point and interval summaries for the static parameters are reported
in Section \ref{ch6ch7:est} for the full model $\mathcal{M}_{F}$ only, with
results pertaining to all competing models recorded in Table
\ref{models}
provided in the on-line supplementary appendix.%

\begin{figure}[ptb]%
\centering
\includegraphics[
natheight=3.133200in,
natwidth=6.530200in,
height=3.1332in,
width=6.5302in
]%
{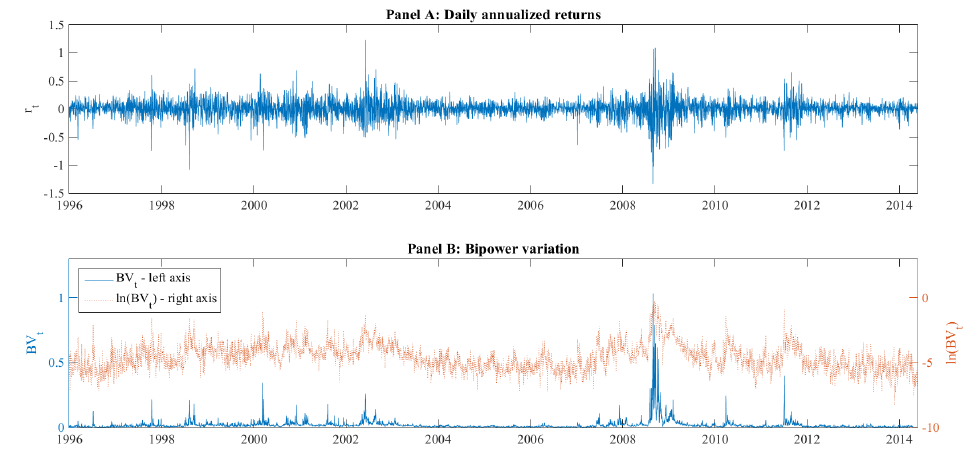}%
\caption{Plots of the S\&P500 logarithmic returns $\left(  r_{t}\right)  $
(Panel A); bipower variation $\left(  BV_{t}\right)  $ and its logarithm
$\left(  \ln BV_{t}\right)  $ (Panel B) for January 3, 1996 to June 23, 2014.}%
\label{r_bv}%
\end{figure}

In Figure \ref{r_bv} we provide a graphical representation of two of the four
measures, $r_{t}$ and $BV_{t}$ (the latter in both raw and logarithmic form)
for the entire sample period, recorded in annualized form. As is evident in
both panels of Figure \ref{r_bv}, and as is completely expected in this
setting, volatility clustering is a marked feature. The most extreme variation
in returns, along with the occurrence of $BV_{t}$ values of unprecedented
magnitude, is observed towards the end of 2008. The large jumps observed
periodically in $BV_{t}$, in addition to the jumps in evidence in the return
series itself, plus the tendency for both types of jumps to cluster, all
provide motivation for the specification of a dynamic model for both price and
volatility jumps.

\begin{figure}[ptb]%
\centering
\includegraphics[
natheight=4.400200in,
natwidth=6.543200in,
height=4.4002in,
width=6.5432in
]%
{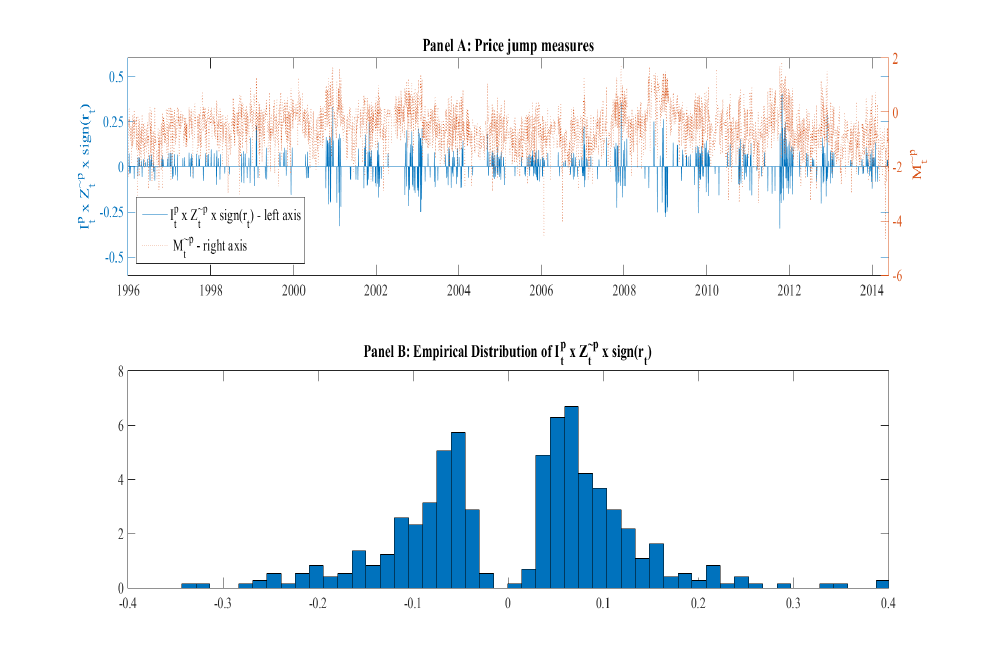}%
\caption{Panel A superimposes two time series plots for January 3, 1996 to
June 23, 2014: i) the solid line (and left-hand-side axis) depicts the product
of the measure of price jump occurrence $\left(  I_{t}^{p}\right)  $, the
measure of price jump size $\left(  \protect\widetilde{Z}_{t}^{p}\right)  $,
and the sign of the return ($r_{t})$; ii) the dotted line (and right-hand-side
axis) depicts the logarithmic price jump size measurement $\left(
\protect\widetilde{M}_{t}^{p}\right)  $. Panel B plots the histogram of the
empirical distribution of the signed price jump measure $\left(
\protect\widetilde{Z}_{t}^{p}\times sign\left(  r_{t}\right)  \right)  $ for
days when the price jump indicator signals the presence of a jump. }%
\label{jump}%
\end{figure}

In Panel A of Figure \ref{jump} we plot the time series of the \textit{signed}
jump size\textbf{ }measure $I_{t}^{p}\times\widetilde{Z}_{t}^{p}\times
sign(r_{t})$, with $I_{t}^{p}$ and $\widetilde{Z}_{t}^{p}$ as defined in
(\ref{ipt}) and (\ref{zp}) respectively, with the data indicating that price
jump intensity is 10.64\% on average.\footnote{With reference to (\ref{ipt}),
$I_{t}^{p}$ is defined using a significance level of $0.001$, as recommended
by Tauchen and\ Zhou (2011).} Values of the combined measure are indicated on
the left-hand-side axis.\footnote{We reiterate that the sign of the price jump
is modelled as a latent process only (in (\ref{s4})), and is estimated along
with all other unknowns in the model. We do not assume that the sign of the
price jump coincides exactly with the sign of the return on that day. We
represent the direction of the price jump by the sign of the return for the
purpose of this preliminary diagnostic exercise only.} Distinct variation in
the observed price jump size over the sample period, including clusterings of
both small and large jumps, is evident, with there being no particular
tendency for \textit{negative}\textbf{ }price jumps (as identified here simply
by the occurrence of a negative return) to predominate over this extended
period. Clusters of large jumps appear intermittently over the sample period;
however, the clusters that are largest in magnitude occur during three of the
most volatile market periods: late 2001 and throughout 2002 following the
September 11 terrorist attacks; the global financial crisis period in 2008 and
2009; and the culmination of the period of Euro-zone debt crises, in 2011. The
logarithmic measure of price jump magnitude, $\widetilde{M}_{t}^{p}$, (as
defined in (\ref{Mpt})), is also included in Panel A, with values indicated on
the right-hand-side axis. The fluctuations in this variable reflect (via the
logarithmic transformation of $\widetilde{Z}_{t}^{p}$) the changes in the
observed price jump size, with changes that are large in magnitude producing
large positive values for $\widetilde{M}_{t}^{p}$, and very small magnitude
changes in $\widetilde{Z}_{t}^{p}$ yielding negative values for $\widetilde{M}%
_{t}^{p}.$%

Panel B of Figure \ref{jump} depicts the histogram of the signed jump size
measure, $I_{t}^{p}\times\widetilde{Z}_{t}^{p}\times sign(r_{t})$. As is
consistent with the time series plot in Panel A, there is no evidence of
negative jumps occurring more frequently than positive jumps throughout the
entire sample period. In addition, the empirical distribution is seen to be
bimodal, with the very small probability mass in the neighbourhood of zero
reflecting the fact that, \textit{conditional} on a significant jump
occurring, the size of that jump is, necessarily, bounded away from
zero.\emph{\textbf{ }}This bimodal feature of the observed price jump
magnitude does not appear to have been recognized in the literature, with a
Gaussian distribution typically adopted for the latent variable $Z_{t}^{p}$.
(See Eraker \textit{et al.}, 2003, and Tauchen and Zhou, 2011, for
example).\emph{ }In contrast, our approach adopts $\widetilde{M}_{t}^{p}$ as a
(noisy) measure of the latent (log) jump size, $M_{t}^{p}$, only when
$\widetilde{Z}_{t}^{p}$ is non-zero, and thereby both accommodates this
observed bimodality and avoids a Gaussian assumption for $Z_{t}^{p}$
itself.\footnote{We are grateful to an anonymous referee who highlighted the
need to accommodate this non-Gaussianity in our modelling of the price jump
size.}

\subsection{The implied Hawkes dynamics\label{ch6ch7:est}}

To illustrate the dynamic structure implied by our full state space model
$\mathcal{M}_{F}$, we provide here posterior summary information relating to
the static parameters, including the parameters of the two jump intensity
processes, corresponding to\textbf{ }the full sample period. Reported in
Table
\ref{ch6BNSest}
are the marginal posterior means (MPMs) and 95\% highest posterior density
(HPD) intervals for the static parameters in (\ref{phi}), calculated
from\ 30,000\ MCMC draws (following a 30,000 draw burn-in period) of which
every 5$^{th}$ draw is saved. Inefficiency factors computed from the retained
posterior draws\ are also reported in the table, estimated as the ratio of the
variance of the sample mean of a set of MCMC draws of a given unknown, to the
variance of the sample mean from a hypothetical independent sample. All
parameter summaries are reported in annualized terms where appropriate. For
example, the magnitude of the parameter $\theta$ accords with an annualized
variance quantity, whilst $\kappa\,$reflects the daily persistence in that
annualized variance. We also record point and interval estimates of the
probability of simultaneous and sequential price and volatility jumps, in the
last two lines in the table.

The inefficiency factors reported in Table
\ref{ch6BNSest}
(for the static parameters) range from 1 to 150, with certain parameters
associated with the variance jump intensity producing the highest values. The
inefficiency factors for all latent variables, computed at selected time
points (and not reported here), range from 3 to 5. The acceptance rates for
all parameters drawn using MH schemes\ range from 15-30\%, with the acceptance
rate for drawing\ $V_{1:T}$ (in blocks) - computed as the proportion of times
that at least one block of $V_{1:T}$\ is updated\ over the entire MCMC chain -
being approximately 99\%. The convergence of the MCMC chains for all unknowns
is also confirmed via inspection of graphical CUSUM plots (Yu and Mykland,
1998), and using the convergence diagnostics prescribed by Heidelberger and
Welch (1983) and Geweke (1992).%

\begin{table}[tbhp] \centering
\caption
{Empirical results for the S\&P 500 stock index for January 3, 1996 to June 23, 2014, inclusive, for the full state space model, $\mathcal
{M}_{F}$.}
\label{ch6BNSest}%
%

\begin{tabular}
[c]{lccc}
&  &  & \\
Parameter & MPM & 95\%\ HPD interval & Inefficiency Factor\\\hline\hline
$\mu$ & 0.199 & (0.139,0.256) & 1.59\\
$\gamma$ & -8.628 & (-9.955,-5.679) & 1.10\\
$\rho$ & -0.357 & (-0.421,-0.289) & 6.54\\\hline
$\mu_{p}$ & -0.419 & (-0.435,-0.403) & 6.43\\
$\gamma_{p}$ & 10.955 & (9.967,11.970) & 22.83\\
$\sigma_{p}$ & 0.207 & (0.187,0.226) & 13.84\\
$\pi_{p}$ & 0.382 & (0.297,0.470) & 12.76\\\hline
$\alpha$ & 8.99$e^{-4}$ & (2.39$e^{-5}$,3.33$e^{-3}$) & 1.82\\
$\beta$ & 0.814 & (0.633,0.956) & 17.17\\
$\sigma_{M_{p}}$ & 0.183 & (0.162,0.203) & 13.47\\\hline
$\psi_{0}$ & 0.970 & (0.796,1.142) & 116.60\\
$\psi_{1}$ & 1.290 & (1.255,1.325) & 81.16\\
$\sigma_{BV}$ & 0.436 & (0.423,0.450) & 5.75\\\hline
$\kappa$ & 0.116 & (0.092,0.167) & 45.97\\
$\theta$ & 8.19$e^{-3}$ & (7.41$e^{-3}$,9.11$e^{-3}$) & 15.79\\
$\sigma_{v}$ & 0.016 & (0.014,0.017) & 19.25\\
$\mu_{v}$ & 9.66$e^{-3}$ & (8.21$e^{-3}$,0.012) & 48.48\\\hline
$\delta_{0}^{p}$ & 0.132 & (0.108,0.170) & 14.96\\
$\alpha_{p}$ & 0.097 & (0.072,0.127) & 9.18\\
$\beta_{pp}$ & 0.062 & (0.047,0.079) & 12.12\\\hline
$\delta_{0}^{v}$ & 0.121 & (0.082,0.158) & 41.23\\
$\alpha_{v}$ & 0.035 & (0.024,0.050) & 149.68\\
$\beta_{vv}$ & 0.030 & (0.021,0.043) & 134.58\\
$\beta_{vp}$ & 5.51$e^{-4}$ & (1.33$e^{-5}$,2.00$e^{-3}$) & 1.77\\
$\beta_{vp}^{\left(  -\right)  }$ & 1.14$e^{-3}$ & (3.11$e^{-5}$,3.84$e^{-3}%
$) & 2.34\\\hline
$\Pr\left(  \Delta N_{t}^{v}=1|\Delta N_{t}^{p}=1\right)  $ & 0.097 &
(0.059,0.139) & 20.35\\
$\Pr\left(  \Delta N_{t+1}^{v}=1|\Delta N_{t}^{p}=1\right)  $ & 0.107 &
(0.066,0.149) & 30.12\\\hline\hline
\end{tabular}
%

\end{table}%

The parameters associated with the two jump intensity processes are, of
course, our primary interest. The\textbf{ }dynamic price jump
intensity,\ $\delta_{t}^{p}$,\ possesses a\ reasonably strong degree of
persistence, as indicated by the relatively low MPM of $\alpha_{p}$,\ and an
$95\%$ HPD interval for $\beta_{pp}$ that is well above zero, consistent with
the presence of self-excitation. The magnitudes of $\alpha_{p}$ and
$\beta_{pp}$ reported here, once annualized, are consistent with the
parameters reported by A\"{\i}t-Sahalia \textit{et al.\ }(2015), who (as noted
earlier) propose a Hawkes process for price jumps, but omit variance jumps in
their stochastic volatility specification.

The MPM of the long-run variance jump intensity, $\delta_{0}^{v}$, is
relatively high compared with previously reported (comparable)\textbf{
}quantities (Eraker \textit{et al.,} 2003, Eraker, 2004 and Broadie \textit{et
al.,} 2007). The variance jump intensity process\emph{\ }is also\textbf{\ }%
more persistent than the price jump intensity process, with the MPM of
$\alpha_{v}$ being lower in magnitude than that of $\alpha_{p}.$ In addition
there is evidence of self-exciting dynamics, as indicated by the\ non-zero MPM
of $\beta_{vv}$. The self-exciting dynamics in $\delta_{t}^{v}$, measured by
$\beta_{vv}$, are much stronger than the feedback from the previous price jump
occurrence, measured by $\beta_{vp}$, and its threshold component, measured by
$\beta_{vp}^{\left(  -\right)  },$ with the marginal posterior densities for
both $\beta_{vp}$ and $\beta_{vp}^{\left(  -\right)  }$ being highly
concentrated around mean values close to zero. The probability of
instantaneous co-jumps, measured by the MCMC-based estimate of $\Pr\left(
\Delta N_{t}^{v}=1|\Delta N_{t}^{p}=1\right)  $, is 9.7\%, whilst the
probability that a volatility jump will follow in the period subsequent to a
price jump is 10.7\%. Thus, whilst the estimated model discounts the
importance of feedback from observed price jumps to volatility jump intensity,
it remains flexible enough to capture the phenomenon of both simultaneous -
and close to simultaneous - price and volatility jumps, with such events
estimated to happen with nearly 20\% probability. Further assessment of the
importance of the dynamic structures specified for price and variance jumps,
and of the presence of jumps \textit{per se}, is conducted in Section
\ref{ch6ch7:eval}, via a comparison of marginal likelihoods.

It is interesting to note that the value of $\kappa$ is rather high compared
to other estimates reported in the literature, with a possible explanation
being that the\ degree of persistence in the latent variance process is
partially captured by the dynamic model for the variance jump intensity in our
specification\footnote{This observation is further supported by the posterior
results (recorded in the on-line supplementary appendix) for the alternative
models listed in Table \ref{models}\emph{.} In brief, diffusive volatility
under those specifications with restrictive assumptions about the dynamics in
volatility jumps ($\mathcal{M}_{3},\mathcal{M}_{4},\mathcal{M}_{8}$ and
$\mathcal{M}_{9}$) is more persistent than otherwise. The unconditional
diffusive variance is also larger in magnitude in these cases.}.
The\textbf{\ }MPM of the other parameters associated with stochastic
volatility, for examples $\rho$, $\sigma_{v}$, $\theta$ and $\mu_{v}$, are
broadly consistent with those reported in the literature (see, for example,
Broadie \textit{et al., }2007 Maneesoonthorn \textit{et al.,} 2012 and
A\"{\i}t-Sahalia, Fan and Li,\ 2013), albeit differing slightly in magnitude
presumably due to the varying sample periods.%

\begin{figure}[ptb]%
\centering
\includegraphics[
natheight=3.199600in,
natwidth=6.230800in,
height=3.1996in,
width=6.2308in
]%
{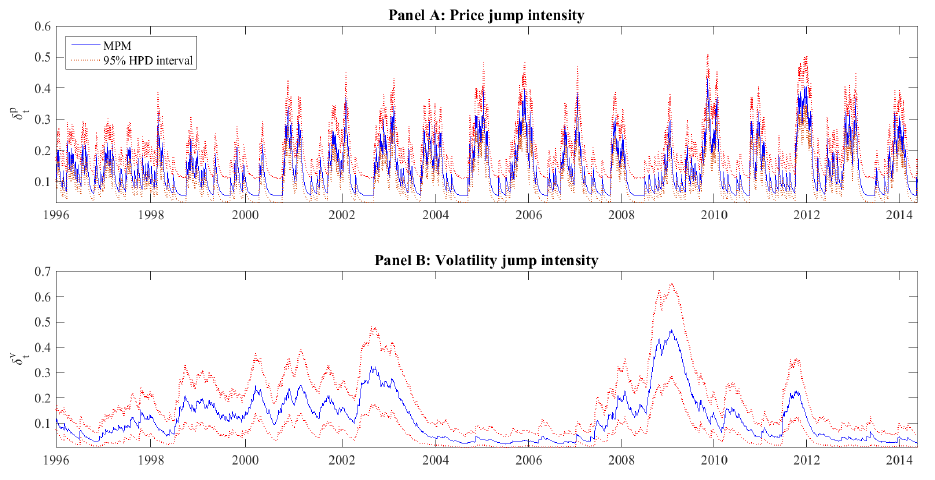}%
\caption{Posterior results for the price jump intensity, $\delta_{t}^{p},$
(Panel\ A) and volatility jump intensity, $\delta_{t}^{v},$ (Panel\ B) over
the period of January 3, 1996 to June 23, 2014. The solid blue lines represent
the marginal posterior means (MPM), while the 95\% HPD intervals are depicted
by the dotted red lines.}%
\label{intensity}%
\end{figure}

Time series plots of the MPMs and the 95\% HPD\ intervals of both jump
intensity processes, $\delta_{t}^{p}$ and $\delta_{t}^{v}$, computed at every
time point over the estimation period, are displayed in Panels A and B,
respectively, of Figure \ref{intensity}.\textbf{ }As is evident from a
comparison of the two panels, the dynamics of the price and volatility jumps
are quite distinct. Price jump clustering - associated with sustained periods
of high values for $\delta_{t}^{p}$ - occurs intermittently throughout the
sample period, and without any obvious tracking of market conditions. An
increase in the intensity of price jumps is both relatively
short-lived\ (compared to that of variance jumps) and associated with periods
in which the magnitude of the observed jumps (Figure \ref{jump}, Panel A) is
either large or small. That is, an increase in price jump intensity does not
appear to\ correlate\ with a period of large price jumps only. The magnitude
of price jumps, however, is found to be associated with the level of
volatility, with the MPM and 95\%\ HPD interval of the parameter $\gamma_{p}$
being in the highly positive region.

\begin{figure}[ptb]%
\centering
\includegraphics[
natheight=2.791600in,
natwidth=5.587600in,
height=2.7916in,
width=5.5876in
]%
{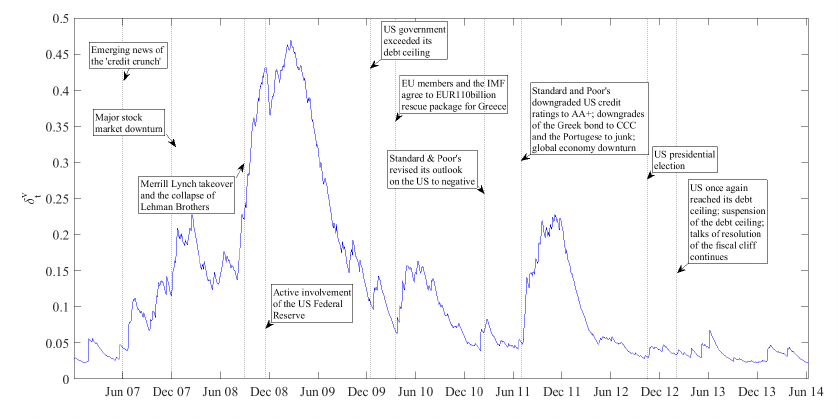}%
\caption{Time series plot of the variance jump intensity process, $\delta
_{t}^{v}$, over a sub-period of January 3, 2007 to June 23, 2014, inclusive,
with the timing of various important market events noted, including the recent
global financial crisis, as well as the events related to the US\ debt ceiling
and Euro-zone debt crises.}%
\label{events}%
\end{figure}

In contrast, the variance\ jumps tend to cluster during high volatility
periods specifically, with an increase in\emph{ }marginal posterior mean and
95\% HPD intervals\textbf{ }associated with $\delta_{t}^{v}$ coinciding with
the rises in the observed volatility measure $BV_{t}$, as recorded in Panel B
of Figure \ref{r_bv}. Some of the sharpest rises in $\delta_{t}^{v}$ are
either synchronous with, or occur soon after, certain key events, as
illustrated in Figure \ref{events}, in which the MPM of $\delta_{t}^{v}$ is
plotted over the 2007-2014 \ period. In particular, the collapse of the Lehman
Brothers (September, 2008)\ and the subsequent intervention by the US Federal
Reserve (December, 2008) are followed closely by the largest variance jump
intensity levels observed throughout the entire sample period (the MPM
reaching a peak of 47\% on\ March 18th, 2009). During the various phases of
the recent US debt ceiling concerns and the Euro-zone debt crisis (starting
from\textbf{\ }late 2009), sharp increases in the MPM of $\delta_{t}^{v}$ are
also\textbf{\ }evident, albeit with the magnitude of these being less than the
rises observed during the global financial crisis. Once a period of multiple
variance\textbf{\ }jumps has passed, the value of $\delta_{t}^{v}$ declines
rather slowly, with this high level of persistence being consistent with the
point and interval estimates of $\alpha_{v}$ recorded in Table
\ref{ch6BNSest}%
.\footnote{The dynamics of volatility jump intensity implied by models
$\mathcal{M}_{5}$ to $\mathcal{M}_{7}$ are not dissimilar to those presented
here, as all three models assume that the jump intensity is driven by the
latent volatility process. The key difference is in the dynamics of the price
jump intensity, with the MPMs and 95\% HPD intervals of $\delta_{t}^{p}$
implied by these three models (and as reported in the on-line supplementary
document) indicating that the price jump intensity is roughly constant. Such a
model-implied feature is obviously inconsistent with the empirical
characteristics of the price jump indicator evident in Figure \ref{jump}
(Panel A).}%

\subsection{Model ranking\label{ch6ch7:eval}}

Table
\ref{BF}
reports the log marginal likelihood of each of the eleven models,
$\mathcal{M}_{1}$ to $\mathcal{M}_{11}$, as well as that of the full model
$\mathcal{M}_{F}$, and as computed over the entire sample period. The Bayes
factor for each of $\mathcal{M}_{1}$ to $\mathcal{M}_{11}$ relative to
$\mathcal{M}_{F}$ are also computed, as per (\ref{ch6bf_if}), as the ratio of
the marginal likelihood of $\mathcal{M}_{F}$ to that of $\mathcal{M}_{i}$,
$i=1,2,...,11$, and are recorded in logarithmic form. We also report the
ranking (from one to twelve) of all of these models, as based on their
marginal likelihoods values.\footnote{As noted in Section \ref{bayes}, a
series of auxiliary MCMC algorithms is required to compute any given Bayes
factor, in addition to the full MCMC algorithm associated with the two models
in question. All auxiliary algorithms produce 10,000 draws, after a 10,000
draw burn-in period.}\emph{ }As noted earlier, the marginal likelihoods of
$\mathcal{M}_{9}$ to\textbf{ }$\mathcal{M}_{11}$, are directly comparable to
those of the other models only if an extra component (based on the two jump
measures) is used to supplement the marginal likelihoods computed directly
from the $r_{t}$ and $BV_{t}$ measures. These augmented figures are recorded
in the middle panel of Table
\ref{BF}%
. For completeness, we also record in the bottom panel of the table the
marginal likelihood based on the $r_{t}$ and $BV_{t}$ measures only, with
these figures not allowing for a direct comparison with the remaining nine models.

The key message from the results recorded in Table
\ref{BF}
is that the\textbf{ }proposed Hawkes specification for both price and
volatility jumps is strongly supported by the data. The log marginal
likelihood of the full dynamic model $\mathcal{M}_{F}$ is only inferior when
compared against its slightly more restrictive alternatives, $\mathcal{M}_{1}$
and $\mathcal{M}_{2}$, which assume no threshold effect and no feedback effect
from price to volatility jumps, respectively. This support for $\mathcal{M}%
_{1}$ and $\mathcal{M}_{2}$ is consistent with the fact that most of the
posterior mass associated with each of $\beta_{vp}$\ and $\beta_{vp}^{\left(
-\right)  }\ $is near zero\textbf{ }in the full dynamic model, $\mathcal{M}%
_{F}$, as indicated by the MPM and 95\% HPD intervals\textbf{ }reported in
Table
\ref{ch6BNSest}%
. The model that\emph{ }\textit{imposes} contemporaneous price and variance
jumps ($\mathcal{M}_{3}$) performs poorly, with the model ranked ninth
overall, indeed ranked more lowly than the model in which no volatility jumps
at all are allowed ($\mathcal{M}_{4}$, ranked sixth) and the model in which
jumps have a constant intensity ($\mathcal{M}_{8}$, ranked eight). All three
models that avoid the Hawkes structure in modelling the dynamic intensities
($\mathcal{M}_{5}$, $\mathcal{M}_{6}$ and $\mathcal{M}_{7}$) are ranked below
both $\mathcal{M}_{F}$ and its two closest restricted versions, $\mathcal{M}%
_{1}$ and $\mathcal{M}_{2}$, and the Heston and RGARCH models ($\mathcal{M}%
_{9}$, $\mathcal{M}_{10}$ and $\mathcal{M}_{11}$) are the most poorly
performing models of all. Of the latter three, when considered in isolation
from the remaining models, the logarithmic RGARCH specification ranks the
highest but does not provide an explanation of the sample data that is close
to any of the models that accommodate jumps.%

\begin{table}[tbhp] \centering
\caption
{Log marginal likelihoods and model rankings, computed using the data from January 3, 1996 to June 23, 2014, inclusive.}
\label{BF}%
%

\begin{tabular}
[c]{ccccc}
&  &  &  & \\
& Model & $\ln\left(  \text{marginal likelihood}\right)  $ & $\ln BF_{i}$ &
Ranking\\\hline\hline
& $\mathcal{M}_{F}$ & -10024 & 0 & 3\\
& $\mathcal{M}_{1}$ & -9861 & $-163$ & 1\\
& $\mathcal{M}_{2}$ & -9957 & $-67$ & 2\\
& $\mathcal{M}_{3}$ & -12868 & $2844$ & 9\\
& $\mathcal{M}_{4}$ & -10639 & $615$ & 6\\
& $\mathcal{M}_{5}$ & -10686 & $662$ & 7\\
& $\mathcal{M}_{6}$ & -10618 & $594$ & 5\\
& $\mathcal{M}_{7}$ & -10076 & $52$ & 4\\
& $\mathcal{M}_{8}$ & -10773 & $749$ & 8\\\hline
with & $\mathcal{M}_{9}$ & -27793 & $17769$ & 10\\
price jump & $\mathcal{M}_{10}$ & -33486 & $23462$ & 12\\
measures & $\mathcal{M}_{11}$ & -26830 & $16806$ & 11\\\hline
without & $\mathcal{M}_{9}$ & -12521 & N/A & N/A\\
price jump & $\mathcal{M}_{10}$ & -18214 & N/A & N/A\\
measures & $\mathcal{M}_{11}$ & -11558 & N/A & N/A\\\hline\hline
\end{tabular}
%

\end{table}%

\subsection{\textbf{Predictive comparison\label{pred_eval}}}

The exercise conducted in the previous section documents the relative
performance of the alternative models over the full sample period. In the
current section, we compute the `joint' CLS in (\ref{cls_full}) and the three
marginal CLS values discussed in Section \ref{pred_perf}, over a more recent
period only, with a training sample used to initialize the computation. Once
again we use the full model $\mathcal{M}_{F}$ as the reference model, but this
time conduct a\textbf{ }comparison of it\textbf{ }only against those
alternative models that are most distinct from it, namely: $\mathcal{M}_{4}$,
in which a Hawkes structure is adopted for price jumps but volatility jumps
are omitted; $\mathcal{M}_{5}$, in which a linear (non-Hawkes) dynamic
structure is adopted for the intensities;\footnote{The relative predictive
performances of models $\mathcal{M}_{6}$ and $\mathcal{M}_{7}$, in which
non-linear functions of $V_{t}$ were used for the jump intensities, were very
similar to that of $\mathcal{M}_{5}.$} $\mathcal{M}_{8}$, in which the jump
intensities are constant; $\mathcal{M}_{9}$, in which no jumps at all are
modelled within the state space structure; and $\mathcal{M}_{10}$ and
$\mathcal{M}_{11}$, which adopt conditionally deterministic specifications for
the variance and also eschew jumps.

The first\thinspace$T_{0}=2500$ observations in the data set are used to
produce, for each model considered, the initial predictive distributions (for
$T_{0}+1$) in both (\ref{cls_full}) and (\ref{marg_cls}). To reduce the
computational burden in obtaining all subsequent predictive distributions, the
posterior distributions for the relevant collection of static parameters are
updated\textbf{ }only every 250 observations thereafter.\textbf{ }For the
state space models, draws of the one-step-ahead latent vector, $X_{t+1}$, are
produced recursively for each of the 2098 trading days, from February 22, 2006
to June 23, 2014, of which the evaluation period is comprised. A particle
filtering algorithm is adopted for this purpose, conditional on the draws of
the static parameters. The candidate state particles are sampled from the
relevant state transition density as the proposal, with the latter being
prescribed by the model in Section \ref{full_SSM} and the restrictions
detailed in Table 1. The predictive ability of the four models under
investigation is evaluated in two ways: in terms of the accuracy of the
probabilistic forecasts of all relevant measurements, assessed by the joint
and marginal cumulative log scores; and in terms of the accuracy of highest
posterior predictive (HPP) interval coverage and Value at Risk (VaR)
prediction for the return measurement alone.

\subsubsection{Cumulative log score assessment \label{Section:CLS}}

\begin{landscape}%

%

\begin{figure}[ptb]%
\centering
\includegraphics[
natheight=4.337900in,
natwidth=9.336500in,
height=4.3379in,
width=9.3365in
]%
{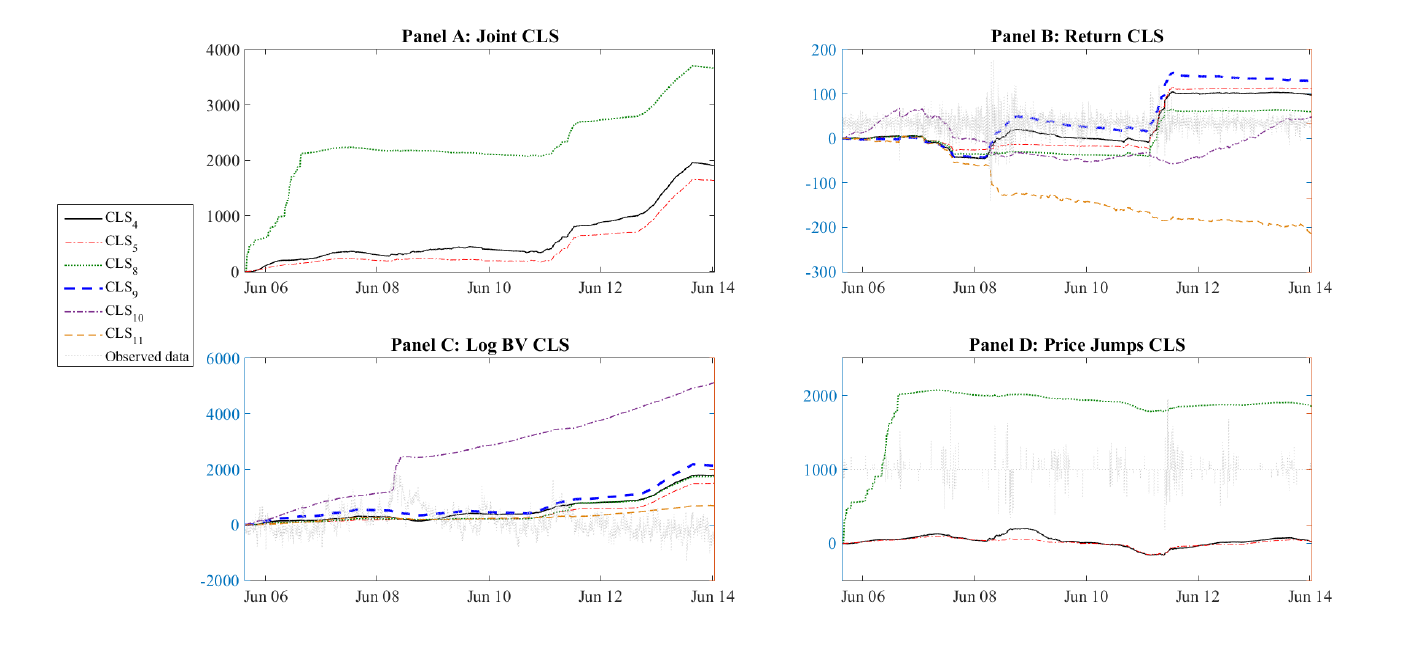}%
\caption{$CLS_{i}$ of $\mathcal{M}_{F}$ relative to\textbf{ }competing models:
$\mathcal{M}_{4}$ (solid black lines), $\mathcal{M}_{5}$ (dash-dot red lines),
$\mathcal{M}_{8}$ (dotted green lines), $\mathcal{M}_{9}$ (thick dashed blue
lines), $\mathcal{M}_{10}$ (thick dashed-dot purple lines) and $\mathcal{M}%
_{11}$ (dashed orange lines) for February 22, 2006 to June 23, 2014. Panels
A-D depict, in turn: the $CLS$ scores for $Y_{t}$, $r_{t}$, $\ln BV_{t}$, and
the joint price jump measure $g_{t}$ given in (\ref{g_predpricejump}). The
relevant observed data over the assessment period are also plotted in the
background of each of panel B-D in light dotted grey.}%
\label{cls}%
\end{figure}
%

\end{landscape}

Panels A to D in Figure \ref{cls} depict, in turn, the joint $CLS$ score
associated with the full measurement vector $Y_{t}$, and the marginal $[g]$
$CLS$ scores of $g_{t}=r_{t}$, $g_{t}=\ln BV_{t}$ and $g_{t}=\left(
\widetilde{M}_{t}^{p},I_{t}^{p}\right)  ^{\prime}$, as given in
(\ref{g_predpricejump}). From Panel A it is clear that the full dynamic model,
$\mathcal{M}_{F}$, dominates all three of the models that exploit the full set
of measurements, $\mathcal{M}_{4}$, $\mathcal{M}_{5}$ and $\mathcal{M}_{8}$,
over the assessment period. The positive CLS scores throughout are consistent
with positive log Bayes factors recorded for the full sample period in Table
\ref{BF}. The results in Panel C are very much in line with those in Panel A,
with $\mathcal{M}_{F}$\ continuing to dominate the comparator models (now
expanded to include $\mathcal{M}_{9}$ to $\mathcal{M}_{11}$) in terms of the
accuracy with which it predicts $\ln BV_{t}$ alone. Somewhat in contrast with
these two sets of results, in Panels B and D the relative performance of
$\mathcal{M}_{F}$ in predicting returns and price jumps respectively is seen
to fluctuate throughout the evaluation period, with $\mathcal{M}_{F}$
sometimes being dominated by certain alternative specifications, despite still
being the best model overall (as indicated by positive final values for both
$CLS$ scores). It is interesting to note (in Panel B) that in terms of
predicting returns, $\mathcal{M}_{F}$ performs the best, amongst all of the
state space models, during high volatility periods - both over the depth of
the GFC in the second half of 2008, and during the Euro-zone debt crisis in
2011 - with all four $CLS_{i}$ curves seen to have strong positive slopes at
those points. Clearly the dynamic specifications incorporated in
$\mathcal{M}_{F}$ have particular predictive power (for returns) during these
turbulent periods. When compared to the conditionally deterministic RGARCH
specifications, $\mathcal{M}_{10}$ and $\mathcal{M}_{11}$, the full state
space model outperforms the linear specification $\mathcal{M}_{10}$ overall,
but under-performs relative to the log-linear specification, $\mathcal{M}%
_{11}$. In predicting the measures related to price jumps alone (Panel\ D),
$\mathcal{M}_{F}$ performs roughly on par with $\mathcal{M}_{4}$ and
$\mathcal{M}_{5}$, both of which employ some sort of dynamic structure for
price jump intensity. However, when compared to $\mathcal{M}_{8}$, the model
with constant jump intensity, $\mathcal{M}_{F}$ clearly dominates.

\subsubsection{Value at risk prediction and HPP coverage}

As a final exercise, we assess the ability of the five alternative models
entertained in Section \ref{Section:CLS} both to accurately estimate
predictive tail quantiles and to produce 95\% HPP intervals with accurate
empirical coverage. We focus here only on the predictive distribution for the
return, with the quantile estimation coinciding with the prediction of 1\% and
5\% VaRs. The empirical coverage statistics associated with both the VaRs and
the HPP intervals, for all five competing models, are reported Table
\ref{tab:CCtest}%
. We also report the results of the Christoffersen (1998) tests of correct
unconditional coverage and independence of exceedances of the (predicted)
intervals. Models that produce forecasts that fail to reject both of these
tests are deemed adequate in predicting VaR.

The results indicate that all seven models being assessed have empirical
coverage that is significantly different from the nominal coverage of the 95\%
HPP intervals over the assessment period. That said, the coverages are all
quite reasonable (in an absolute sense) and $\mathcal{M}_{F}$ performs on par
with $\mathcal{M}_{11}$, with empirical coverages that are quite close to the
95\% level, as well as being the only models that do not reject the null
hypothesis of independent violations. In all but two cases - the 1\% VaR
prediction from $\mathcal{M}_{5}$ and the 5\% VaR prediction from
$\mathcal{M}_{10}$ - the competing models produce VaR predictions with
independent exceedences, with $\mathcal{M}_{F}$ being one of the three models
with the empirical tail coverage closest to the nominal quantile
probabilities. Perhaps not surprisingly, the worst performance (in terms of
both HPP and tail coverage) is exhibited by the (Heston) model, $\mathcal{M}%
_{9}$, in which no price or volatility jump components feature.

In summary then, these results are consistent with the rankings produced by
the marginal CLS computations for the return, as reported in the previous
section. They confirm the importance of including both price and variance
jumps in this empirical setting and, moreover, highlight the added value of
augmenting the basic stochastic volatility structure with the particular
dynamic structure for the price and volatility jumps as represented by a
Hawkes process.%

%

\begin{table}[tbhp] \centering
\caption
{Empirical tail coverage, computed as the proportion of observed returns that are lower than the 5\% and 1\% VaR
predictions, respectively, is given in Column 2 and 3. The empirical coverage of the 95\% HPP interval of the
predictive returns distribution is given in Column 4. The superscripts * and + denote empirical coverage that is
statistically different from the nominal level, and whose exceedences fail the independence test at the 5\%
significance level, respectively. All statistics are computed over the assessment period of February 22, 2006 to June 23, 2014, inclusive. }
\label{tab:CCtest}%
%

\begin{tabular}
[c]{cccc}
&  &  & \\
& \multicolumn{2}{c}{Empirical tail coverage} & \multicolumn{1}{|c}{Empirical
coverage}\\
& 5\% VaR & 1\% VaR & \multicolumn{1}{|c}{95\% HPP interval}\\\hline\hline
$\mathcal{M}_{F}$ & 7.34\%$^{\ast}$ & 2.86\%$^{\ast}$ &
\multicolumn{1}{|c}{91.94\%$^{\ast}$}\\
$\mathcal{M}_{4}$ & 8.58\%$^{\ast}$ & 3.91\%$^{\ast}$ &
\multicolumn{1}{|c}{90.18\%$^{\ast+}$}\\
$\mathcal{M}_{5}$ & 8.06\%$^{\ast}$ & 3.05\%$^{\ast+}$ &
\multicolumn{1}{|c}{90.75\%$^{\ast+}$}\\
$\mathcal{M}_{8}$ & 7.96\%$^{\ast}$ & 2.81\%$^{\ast}$ &
\multicolumn{1}{|c}{91.09\%$^{\ast+}$}\\
$\mathcal{M}_{9}$ & 9.01\%$^{\ast}$ & 4.62\%$^{\ast}$ &
\multicolumn{1}{|c}{89.37\%$^{\ast+}$}\\
$\mathcal{M}_{10}$ & 4.62\%$^{+}$ & 1.67\%$^{\ast}$ &
\multicolumn{1}{|c}{96.38\%$^{\ast+}$}\\
$\mathcal{M}_{11}$ & 8.91\%$^{\ast}$ & 4.36\%$^{\ast}$ &
\multicolumn{1}{|c}{91.28\%$^{\ast}$}\\\hline\hline
\end{tabular}
%

\end{table}%

\section{Conclusions\label{ch6concl7}}

In this paper a very flexible stochastic volatility model is proposed, in
which dynamic behaviour in price and variance\ (and, hence, volatility) jumps
is accommodated via a bivariate Hawkes process for the two jump intensities.
The model allows both price and variance jumps to cluster over time, for the
two types of jump to occur simultaneously, or otherwise, and for the
occurrence of a price jump to impact on the likelihood of a subsequent
variance jump. A nonlinear state space model that uses daily returns on the
S\&P500 market index, in addition to nonparametric measures of volatility and
price jumps, is constructed, with a hybrid Gibbs-MH MCMC algorithm used to
estimate the model and compute marginal likelihoods and various predictive
quantities. As remains standard in the literature, given that within-day index
data informs the analysis, the conclusions we draw regarding the dynamics in
asset prices pertain to within-day movements only, with the inclusion of
overnight movements potentially requiring a modified set of assumptions to be
adopted regarding the factors driving the dynamics therein.

A large number of alternative models, many of which impose restrictions on the
general state space specification, are explored using Bayes factors, with the
overall conclusion being in favour of the models that specify Hawkes dynamics
in both price and variance jump\ intensity. Based on the most general
specification, the probability of price and volatility jumps occurring either
on the same day or on successive days is estimated to be close to 20\% and the
price jump size is found to be associated with the latent volatility itself.
The dynamic structures imposed on the occurrences of price and variance jumps
are also shown to add value to the predictions of returns on the index
(including VaR predictions), as well as to the prediction of the nonparametric
measures of volatility and jumps. One particular (conditionally deterministic)
alternative - the logarithmic form of RGARCH\ - performs the best of all
models in terms of the CLS for the return, but does not dominate the more
complex state space specifications in terms of predicting (logarithmic)
bipower variation, and is unable to be used to predict jumps of any sort.

Perhaps not surprisingly, our investigation suggests that the price jump
intensity possesses qualitatively different time series behaviour from that of
the variance jump intensity.\ Clusters of inflated price jump intensities are
relatively short-lived and scattered throughout the sample period, whilst
clusters of high variance jump intensities\ occur less frequently but persist
for longer when they do occur. Furthermore, rises in the intensity of variance
jumps are very closely associated with negative market events, whereas as no
corresponding link is evident for the price jump intensity.

Having thus quantified the importance of dynamic jumps - and of respecting the
particular nature of the interaction between price and volatility jumps - in
the modelling of index returns, such features would appear to deserve more
careful attention in future risk management strategies. Importantly though,
further work is also required to ascertain the robustness of our qualitative
results to the manner in which high frequency data is used to measure the
occurrence and size of jumps (see Dumitru and Urga, 2012) and to the use of
observed quarticity measures in the modelling of integrated variance
\textbf{(}see, for example, Dobrev and Szerszen, 2010, and Bollerslev, Patton
and Quaedvlieg, 2016). Extensive work along these lines is currently being
undertaken by the authors.

\paragraph{Appendix A: Prior specification}

\baselineskip14pt

Uniform priors are assumed for the parameters $\kappa$ and $\theta,$ truncated
from below at zero,\textbf{ }while the parameter $\sigma_{v}^{2}$ is blocked
with the leverage parameter, $\rho$, via the reparameterization: $\psi
=\rho\sigma_{v}$ and $\omega=\sigma_{v}^{2}-\psi^{2}$; see Jacquier, Polson
and Rossi (2004). This reparameterization is convenient as, given $V_{1:T}$,
it allows $\psi$ and $\omega$ to be treated respectively as the slope and
error variance coefficients in a normal linear regression model. Direct
sampling of $\psi$ and $\omega$ is then conducted using standard posterior
results, based on conjugate prior specifications in the form of
conditional\ normal and inverse gamma distributions, respectively, given by
$p\left(  \psi|\omega\right)  \sim N\left(  \psi_{0}=-0.005,\sigma_{0}%
^{2}=\omega/5.0\right)  $ and $p\left(  \omega\right)  \sim IG\left(
a=10,b=0.001\right)  $, where $b$\ denotes the scale parameter in the context
of the inverse gamma distributions discussed here. The prior specifications
for $\psi$ and $\omega$ are chosen such that the implied prior distributions
for $\rho$ and $\sigma_{v}$ are relatively diffuse, with the ranges being
broadly in line with the range of the empirical values of these parameters
reported in the literature.\ 

Truncated uniform priors are specified for the parameters $\mu,$ $\gamma,$
$\mu_{p}$ and $\gamma_{p}$. Very wide ranges of values for these parameters,
over both the negative and positive regions of the real line, are thus
specified \textit{a priori}. The volatility feedback parameter $\gamma$ is
assumed \textit{a priori} to be bounded from above at zero, which is
consistent with recent findings of negative volatility feedback in the high
frequency literature. (See, for example, Bollerslev \textit{et al.} 2006, and
Jensen and\ Maheu 2014). Conjugate inverse gamma priors are applied to the
parameters $\sigma_{p}^{2}$ and $\sigma_{BV}^{2}$, with both prior
distributions being centred around a mean of 0.5, and with (a relatively
large) standard deviation of 0.5.

Conjugate beta priors are employed for the unconditional jump intensities,
$\delta_{0}^{p}$ and $\delta_{0}^{v}.$ The hyperparameters of these priors are
chosen such that the prior mean of 0.1 matches the sample mean of the observed
$\Delta N_{1:T}^{p}$. The prior distribution of $\delta_{0}^{v}$ is, in turn,
equated with that of $\delta_{0}^{p}$, stemming from the prior belief that if
there is a price jump ($\Delta N_{t}^{p}=1$), then it is \textit{likely}
(albeit not strictly necessary) that the variance process also contains a jump
(that is, $\Delta N_{t}^{v}=1$). A conjugate inverse gamma prior is employed
for $\mu_{v}$, implying a prior mean of 0.007 and prior standard deviation of
0.002, where this prior mean is a proportion of the average of $\max\left(
RV_{t}-BV_{t},0\right)  .$ The initial stochastic variance is assumed to be
degenerate, with $V_{1}=\theta+\frac{\mu_{v}\delta_{0}^{v}}{\kappa}$. Uniform
priors are employed for the jump intensity parameters, $\alpha_{p},\beta
_{pp},\beta_{vp},\beta_{vp}^{\left(  -\right)  },\alpha_{v}$ and $\beta_{vv}$,
conforming to the theoretical restrictions listed in Section
\ref{Section:model}, and the prior belief that $\beta_{vp}>0$ and $\beta
_{vp}^{\left(  -\right)  }>0.$ The prior mean and standard deviation for each
parameter is documented in Table \ref{priors}.%

\begin{table}[tbhp] \centering
\caption
{Prior specifications for each of the elements of the parameter vector $\phi$}
\label{priors}%
%

\begin{tabular}
[c]{cccc}
&  &  & \\
Parameter & Prior Spec & Mean & Stdev\\\hline\hline
$\mu$ & $U\left(  -10,10\right)  $ & $0$ & $5.77$\\
$\gamma$ & $U\left(  -10,0\right)  $ & $-5$ & $2.89$\\
$\rho$ & $\rho,\sigma_{v}$ joint & $-0.34$ & $0.33$\\\hline
$\mu_{p}$ & $U\left(  -100,100\right)  $ & $0$ & $57.7$\\
$\gamma_{p}$ & $U\left(  0,100\right)  $ & $50$ & $28.9$\\
$\sigma_{p}^{2}$ & $IG\left(  a=3,b=1\right)  $ & $0.5$ & $0.5$\\
$\pi_{p}$ & $\beta\left(  a=5,b=5\right)  $ & $0.5$ & $0.15$\\
$\alpha$ & $\beta\left(  a=0.01,b=10\right)  $ & $0.001$ & $0.01$\\
$\beta$ & $\beta\left(  a=7,b=3\right)  $ & $0.7$ & $0.14$\\
$\sigma_{M_{p}}^{2}$ & $IG\left(  a=3,b=1\right)  $ & $0.5$ & $0.5$\\\hline
$\psi_{0}$ & $N\left(  0,0.1\right)  $ & $0$ & $0.1$\\
$\psi_{1}$ & $N\left(  1,0.1\right)  $ & $1$ & $0.1$\\
$\sigma_{BV}^{2}$ & $IG\left(  a=3,b=1\right)  $ & $0.5$ & $0.5$\\\hline
$\kappa$ & $U\left(  0,1\right)  $ & $0.5$ & $0.29$\\
$\theta$ & $U\left(  0,0.1\right)  $ & $0.05$ & $0.03$\\
$\sigma_{v}$ & $\rho,\sigma_{v}$ joint & $0.012$ & $0.003$\\
$\mu_{v}$ & $IG\left(  a=20,b=1/7.2\right)  $ & $7e^{-3}$ & $2e^{-3}$\\\hline
$\delta_{0}^{p}$ & $\beta\left(  a=1,b=9\right)  $ & $0.1$ & $0.03$\\
$\alpha_{p}$ & $U\left(  0,1\right)  $ & $0.5$ & $0.29$\\
$\beta_{pp}$ & $U\left(  0,1\right)  $ & $0.5$ & $0.29$\\\hline
$\delta_{0}^{v}$ & $\beta\left(  a=1,b=9\right)  $ & $0.1$ & $0.03$\\
$\alpha_{v}$ & $U\left(  0,1\right)  $ & $0.5$ & $0.29$\\
$\beta_{vv}$ & $U\left(  0,1\right)  $ & $0.5$ & $0.29$\\
$\beta_{vp}$ & $U\left(  0,1\right)  $ & $0.5$ & $0.29$\\
$\beta_{vp}^{\left(  -\right)  }$ & $U\left(  0,1\right)  $ & $0.5$ &
$0.29$\\\hline\hline
\end{tabular}
%

\end{table}%

\paragraph{Appendix B.1: MCMC algorithm for $\mathcal{M}_{F}$}

The MCMC algorithm for sampling from the joint posterior in (\ref{post}) can
be broken down into seven main steps, as outlined below:

\begin{algorithm}
At each iteration:

\begin{enumerate}
\item Sample $V_{1:T}$ in blocks of random length from $V_{1:T}|Z_{1:T}%
^{v},\Delta N_{1:T}^{v},M_{1:T}^{p},\Delta N_{1:T}^{p},S_{1:T}^{Z_{p}}%
,Y_{1:T},\phi$ using MH sampling as described below

\item Sample $\Delta N_{1:T}^{v}$ in a single block from $\Delta N_{1:T}%
^{v}|V_{1:T},Z_{1:T}^{v},M_{1:T}^{p},\Delta N_{1:T}^{p},S_{1:T}^{Z_{p}%
},Y_{1:T},\phi$ using the conditionally independent Bernoulli structure

\item Sample $Z_{1:T}^{v}$\ in a single block from $Z_{1:T}^{v}|V_{1:T},\Delta
N_{1:T}^{v},M_{1:T}^{p},\Delta N_{1:T}^{p},S_{1:T}^{Z_{p}},Y_{1:T},\phi$ using
the conditionally independent truncated normal structure

\item Sample $\Delta N_{1:T}^{p}$ in a single block from $\Delta N_{1:T}%
^{p}|V_{1:T},Z_{1:T}^{v},\Delta N_{1:T}^{v},M_{1:T}^{p},S_{1:T}^{Z_{p}%
},Y_{1:T},\phi$ using the conditionally independent Bernoulli structure

\item Sample $M_{1:T}^{p}$ in a single block from $M_{1:T}^{p}|V_{1:T}%
,Z_{1:T}^{v},\Delta N_{1:T}^{v},\Delta N_{1:T}^{p},S_{1:T}^{Z_{p}}%
,Y_{1:T},\phi$ using the conditionally independent normal structure

\item Sample $S_{1:T}^{Z_{p}}$ in a single block from $S_{1:T}^{Z_{p}}%
|V_{1:T},Z_{1:T}^{v},\Delta N_{1:T}^{v},M_{1:T}^{p},\Delta N_{1:T}^{p}%
,Y_{1:T},\phi$ using the conditionally independent Bernoulli structure

\item Sample $\phi$ from $\phi|X_{1:T},Y_{1:T}$ as described below
\end{enumerate}
\end{algorithm}

The most challenging part of the algorithm is step 1, namely\textbf{ }the
generation of the variance process $V_{1:T}$, due to the nonlinear functions
of $V_{t}$ that feature in the measurement equations (\ref{m1}) and
(\ref{m4}), and in the state equation (\ref{s1}).\ As in Maneesoonthorn
\textit{et al.} (2012) - in which a nonlinear state space model is specified
for both option- and spot-price based measures, and forecasting risk premia is
the primary focus - we adopt a multi-move algorithm for the latent volatility
that extends an approach suggested by\textbf{ }Stroud, M\"{u}ller and
Polson\textbf{ }(2003). In the current context this involves augmenting the
state space model with mixture indicator vectors corresponding to the latent
variance vector $V_{1:T}$ and the two observation vectors $r_{1:T}$ and $\ln
BV_{1:T}$. Conditionally, the mixture indicators define suitable
linearizations of the relevant state or observation equation and are used to
establish a linear Gaussian candidate model for use within an MH subchain.
Candidate vectors of $V_{1:T}$ are sampled and evaluated in blocks. With due
consideration taken of the different model structure and data types, Appendix
A of Maneesoonthorn \textit{et al.} provides sufficient information for the
details of this component of the algorithm applied herein to be extracted.

The elements of $\phi$ are sampled in step 7 using MH subchains wherever
necessary. Given\textbf{ }the draws of $V_{1:T}\ $and $M_{1:T}^{p}$, and all
of the unknowns that appear in (\ref{m1}) - (\ref{d2}), the parameters $\mu$,
$\gamma$, $\mu_{p}$, $\gamma_{p}$, $\psi_{0}$ and $\psi_{1}$ can be treated as
regression coefficients, with exact draws produced in the standard manner from
Gaussian conditional posterior distributions, appropriately truncated as a
consequence of the previously specified\textbf{\ }priors. The sampling schemes
of the conditional variance terms $\sigma_{BV}^{2}$, $\sigma_{M_{p}}^{2}$and
$\sigma_{p}^{2}$ are standard, with inverse gamma conditional posteriors.
Similarly, parameters $\pi_{p}$, $\alpha$ and $\beta$ are sampled using Gibbs
schemes, as all three have closed form conditional beta posteriors. As
described in Appendix A, the parameters $\rho$ and $\sigma_{v}$ are sampled
indirectly via the conditionals of $\psi=\rho\sigma_{v}$ and $\omega
=\sigma_{v}^{2}-\psi^{2}$, which take the form of normal and inverse gamma
distributions, respectively. Conditional upon the draws of $V_{1:T}$, $\Delta
N_{1:T}^{v}$ and $Z_{1:T}^{v}$, the parameters $\kappa$, $\theta$, $\psi$ and
$\omega$ are drawn in blocks, taking advantage of the (conditionally) linear
regression structure with truncated Gaussian errors, and with the constraint
$\sigma_{v}^{2}\leq2\kappa\theta$ imposed.

The static parameters associated with the price and variance jump processes
are dealt with as follows.\textbf{\ }The mean of the variance jump size,
$\mu_{v}$, is sampled directly from an inverse gamma distribution, and the
unconditional jump intensities, $\delta_{0}^{p}$ and $\delta_{0}^{v}$ are
sampled directly from beta posteriors. Each of the parameters, $\alpha
_{p},\beta_{pp},\alpha_{v},\beta_{vv},\beta_{vp},\beta_{vp}^{\left(  -\right)
}$, is sampled using an appropriate candidate beta distribution in an MH
algorithm, subject to restrictions that ensure that (\ref{d1}) and (\ref{d2})
define stationary processes, and that (\ref{delta_p_0}) and (\ref{delta_v_0})
are defined on the $[0,1]$ interval. The intensity parameters $\delta_{\infty
}^{p}$ and $\delta_{\infty}^{v}$, are then computed using the explicit
relationships in (\ref{delta_p_0}) and (\ref{delta_v_0}), and the vectors
$\delta_{1:T}^{v}$ and $\delta_{1:T}^{p}$ updated deterministically based on
(\ref{d1}) and (\ref{d2}).

The algorithms for all comparator state space\textbf{ }models described in
Section \ref{bayes}, $\mathcal{M}_{i}$, for $i=1,...,9$, proceed in an
analogous way.

\paragraph{Appendix B.2: MCMC algorithm for the RGARCH models}

The joint posterior for the RGARCH models $\mathcal{M}_{10}$ and
$\mathcal{M}_{11}$ satisfies $p\left(  \phi\mathbf{|}Y_{1:T}\right)  \propto
p\left(  Y_{1}|\phi\right)  p\left(  \phi\right)  \left[
{\textstyle\prod\limits_{t=2}^{T}}
p\left(  Y_{t}|Y_{1:t-1,}\phi\right)  \right]  ,$\emph{ }with $Y_{t}=\left(
r_{t},BV_{t}\right)  ^{\prime}$ for $\mathcal{M}_{10}$ and $Y_{t}=\left(
r_{t},\ln BV_{t}\right)  ^{\prime}$ for $\mathcal{M}_{11}$. For the purpose of
estimation, we employ the variance targeting approach, and reparameterize
$\omega=\sigma_{0}^{2}\left(  1-\beta-\gamma\right)  $, with $\sigma_{0}^{2}$
denoting the unconditional variance of the return. The elements of the
parameter vector are identical for the two models: $\phi=\left(  \sigma
_{0}^{2},\beta,\gamma,\xi,\varphi,\tau_{1},\tau_{2},\sigma_{u}^{2}\right)
^{\prime}$. We impose noninformative priors on $\phi$: inverse gamma priors,
$IG\left(  a=3,b=1\right)  $, are employed for both $\sigma_{0}^{2}$ and
$\sigma_{u}^{2}$; uniform priors on the unit interval are employed for $\beta$
and $\gamma$; and the priors for $\xi,\varphi,\tau_{1}$ and $\tau_{2}$ are
uniform between $-20$ and $+20$. Since there are no latent variables involved
in the model, the MCMC\ algorithm to sample from the joint posterior is quite
straightforward, with MH\textbf{ }steps required only for $\sigma_{0}^{2},$
$\beta$ and $\gamma$.

\paragraph{Appendix C: Marginal likelihood computation}

The basic idea underlying the evaluation of (\ref{ch6marg_lik}) is the
recognition that it can be re-expressed as
\begin{equation}
p\left(  Y_{1:T}|\mathcal{M}_{i}\right)  =\frac{p\left(  Y_{1:T}|\phi
_{i},\mathcal{M}_{i}\right)  p\left(  \phi_{i}|\mathcal{M}_{i}\right)
}{p\left(  \phi_{i}|Y_{1:T},\mathcal{M}_{i}\right)  }, \label{ch6my}%
\end{equation}
for any point $\phi_{i}$ in the posterior support\ of model\ $\mathcal{M}_{i}%
$, where $\phi_{i}$ denotes the vector of static parameters associated with
model\textbf{ }$\mathcal{M}_{i}$. The first component of the numerator on the
right-hand-side of (\ref{ch6my}) is the likelihood, conditional on
$\mathcal{M}_{i}$, marginal of the latent variables. That is,
\begin{equation}
p\left(  Y_{1:T}|\phi_{i},\mathcal{M}_{i}\right)  =\int p\left(
Y_{1:T}|X_{1:T}^{(i)},\phi_{i},\mathcal{M}_{i}\right)  p\left(  X_{1:T}%
^{(i)}|\phi_{i},\mathcal{M}_{i}\right)  dX_{1:T}^{(i)} \label{mlike}%
\end{equation}
The denominator on the right-hand-side of (\ref{ch6my}) is simply the
conditional posterior density of the (static) parameter vector, also
marginalized\textbf{\ }over the latent variables,%
\begin{equation}
p\left(  \phi_{i}|Y_{1:T},\mathcal{M}_{i}\right)  =\int p\left(  \phi
_{i}|Y_{1:T},X_{1:T}^{(i)},\mathcal{M}_{i}\right)  dX_{1:T}^{(i)}.
\label{ch6post}%
\end{equation}
The evaluation of (\ref{mlike}) at a high density posterior point $\phi
_{i}^{\ast}$ (say, the vector of marginal posterior means for the elements of
$\phi_{i}$) is straightforward, using the output of a full MCMC run for model
$\mathcal{M}_{i}$; namely, the closed form representation of $p\left(
Y_{1:T}|X_{1:T}^{(i)},\phi_{i},\mathcal{M}_{i}\right)  $ is averaged over the
draws of the latent states, $X_{1:T}^{(i)}$, and computed at the given point
$\phi_{i}^{\ast}.$ Evaluation of (\ref{ch6post}) is more difficult, in
particular when a combination of Gibbs and MH algorithms needs to be employed
in the production of draws of $\phi_{i}$. Exploiting the structure of
the\textbf{ }posterior density, we decompose $p\left(  \phi_{i}^{\ast}%
|Y_{1:T},\mathcal{M}_{i}\right)  $\ into five constituent densities as:
\begin{equation}
p\left(  \phi_{i}^{\ast}|Y_{1:T},\mathcal{M}_{i}\right)  =p\left(  \phi
_{1i}^{\ast}|Y_{1:T},\mathcal{M}_{i}\right)  p\left(  \phi_{2i}^{\ast}%
|\phi_{1i}^{\ast},Y_{1:T},\mathcal{M}_{i}\right)  \cdots p\left(  \phi
_{5i}^{\ast}|\phi_{1i}^{\ast},\phi_{2i}^{\ast},...,\phi_{4i}^{\ast}%
,Y_{1:T},\mathcal{M}_{i}\right)  , \label{joint_prod}%
\end{equation}
where $\phi_{1i}=\left(  \sigma_{BV},\mu_{v},\delta_{0}^{p},\delta_{0}%
^{v},\rho,\sigma_{v},\mu_{p},\alpha,\beta,\pi_{p},\sigma_{M_{p}}\right)
,$\ $\phi_{2i}=\left(  \alpha_{p},\alpha_{v},\kappa,\gamma_{p},\psi
_{0}\right)  ,$ $\phi_{3i}=(\beta_{pp},\beta_{vv},\theta,$ $\sigma_{p}%
,\psi_{1}),$\ $\phi_{4i}=\left(  \beta_{vp},\mu\right)  ,$ and $\phi
_{5i}=\left(  \beta_{vp}^{\left(  -\right)  },\gamma\right)  .$ Following the
methods outlined by Chib (1995) and Chib and Jeliazkov (2001), five additional
auxiliary\textbf{ }MCMC chains, each of which involves a different level of
conditioning and, hence, a reduced number of free parameters, are then run to
estimate each of the last five components of (\ref{joint_prod}), in turn
evaluated at $\phi_{ji}^{\ast},$ $j=2,...,5.$\ The first component on the
right hand side of (\ref{joint_prod}), involving no such conditioning, is
estimated from the output of the full MCMC chain, in the usual way.\textbf{ }

Calculation of the marginal likelihoods of the RGARCH models follows
similarly, albeit without the latent variables playing a role, and with the
choice of the auxiliary chains being determined by nature of the parameter
sets for these models. The marginal likelihood for $\mathcal{M}_{10}$ also
includes a Jacobian factor that accounts for the fact that $\mathcal{M}_{10}$
specifies a model for the raw measure $BV_{t}$, whereas all others are
specified in terms of the transformed measure, $\ln BV_{t}$.

Finally,\textbf{ }two versions of the marginal likelihood for models
$\mathcal{M}_{9},$ $\mathcal{M}_{10}$ and $\mathcal{M}_{11}$ are produced: one
that only considers measurements that are directly used in the model, with
$Y_{t}=\left(  r_{t},\ln BV_{t}\right)  ^{\prime}$; and one that employs the
full measurement set,\textbf{ }$Y_{t}=\left(  r_{t},\ln BV_{t},I_{t}%
^{p},\widetilde{M}_{t}^{p}\right)  ^{\prime}$. The second form of marginal
likelihood\textbf{ }allows for the comparison across all models considered in
the paper. Since the possibility of price jumps is actually excluded in each
of\emph{ }$\mathcal{M}_{9},$ $\mathcal{M}_{10}$ and $\mathcal{M}_{11}$, we
employ the specifications: $I_{t}^{p}\sim Bernoulli\left(  \alpha\right)  $
for the price jump occurrence and $\widetilde{M}_{t}^{p}\sim N\left(
-10,\sigma_{M_{p}}^{2}\right)  $ for the log price jump size, with the priors
for $\alpha$ and $\sigma_{M_{p}}^{2}$ defined in Appendix A. The specification
for $I_{t}^{p}$ is nested in (\ref{m2}), associated with $\Delta N_{t}^{p}=0$
for all $t$\emph{. }The prior expectation of $\widetilde{M}_{t}^{p}$ is
assumed to be a large negative value as this reflects a price jump magnitude
that is close to zero. The marginal likelihood components related to these
measures are straightforward to evaluate, with the closed form expressions of
$p\left(  I_{t}^{p}|\mathcal{M}_{j}\right)  $ and $p\left(  \widetilde{M}%
_{t}^{p}|\mathcal{M}_{j}\right)  $ being available analytically for $j=9,10$
and\textbf{ }$11$.

\end{document}


\title{Inference on Self-Exciting Jumps in Prices and Volatility using High Frequency
Measures.\bigskip\\Supplementary Appendix: Results for models\\$\mathcal{M}_{1}$ to $\mathcal{M}_{11}$}
\author{Worapree Maneesoonthorn\thanks{O.Maneesoonthorn@mbs.edu. Melbourne Business
School, The University of Melbourne. }, Catherine S.
Forbes\thanks{Catherine.Forbes@monash.edu. Department of Econometrics and
Business Statistics, Monash University.} and Gael M.
Martin\thanks{Corresponding author: gael.martin@monash.edu. Department of
Econometrics and Business Statistics, Monash University.}}
\maketitle

\begin{abstract}
In this supplement we provide additional posterior results that complement
those documented in Section 4 of the main text. Specifically, we report
Bayesian point and interval estimates of the static parameters of models
$\mathcal{M}_{1}$ to $\mathcal{M}_{11}$ (specified in Table 1 of the main
text). The prior distributions described in Table 5 (Appendix A of the main
text) are employed - where appropriate - for the nested models. These prior
distributions are also applied to the common parameters in the non-nested
models $\mathcal{M}_{5}$ to $\mathcal{M}_{7}$, with the priors for the jump
intensity parameters in those models\ being uniform and conforming to the
theoretical restrictions that the model-implied unconditional jump intensities
are between 0 and 1. The prior distributions employed for the realized GARCH
specifications $\mathcal{M}_{10}$ to $\mathcal{M}_{11}$ conform to the
stationarity conditions underpinning the model.\textbf{\ }All eleven models
are estimated using the S\&P500 data over the sample period from January 3,
1996 to June 23, 2014, inclusive. The marginal posterior means (MPMs), 95\%
highest posterior density (HPD) intervals, along with the inefficiency factors
associated with the relevant MCMC draws are recorded in Tables A1 to A11,
respectively. Each table also contains the model-implied instantaneous and
time lagged co-jump statistics.

\end{abstract}

\newpage

Table A1: Posterior summaries for model $\mathcal{M}_{1}$, based on S\&P500
stock index data from January 3, 1996 to June 23, 2014, inclusive.

\begin{center}%
\begin{tabular}
[c]{lccc}
&  &  & \\
\multicolumn{4}{l}{$\mathcal{M}_{1}$: $\beta_{vp}^{\left(  -\right)  }=0$%
}\\\hline\hline
Parameter & MPM & 95\%\ HPD interval & Inefficiency Factor\\\hline\hline
$\mu$ & 0.197 & (0.137,0.253) & 1.53\\
$\gamma$ & -8.610 & (-9.961,-5.540) & 1.02\\
$\rho$ & -0.355 & (-0.420,-0.286) & 7.35\\\hline
$\mu_{p}$ & -0.412 & (-0.438,-0.405) & 6.70\\
$\gamma_{p}$ & 11.297 & (10.081,12.390) & 30.16\\
$\sigma_{p}$ & 0.208 & (0.189,0.226) & 14.37\\
$\pi_{p}$ & 0.382 & (0.301,0.466) & 11.43\\\hline
$\alpha$ & 9.16$e^{-4}$ & (2.73$e^{-5}$,3.41$e^{-3}$) & 1.83\\
$\beta$ & 0.803 & (0.613,0.948) & 16.21\\
$\sigma_{M_{p}}$ & 0.182 & (0.162,0.202) & 15.34\\\hline
$\psi_{0}$ & 1.044 & (0.772,1.228) & 204.06\\
$\psi_{1}$ & 1.303 & (1.253,1.340) & 144.57\\
$\sigma_{BV}$ & 0.436 & (0.422,0.450) & 6.59\\\hline
$\kappa$ & 0.116 & (0.092,0.142) & 54.74\\
$\theta$ & 8.08$e^{-3}$ & (7.30$e^{-3}$,8.96$e^{-3}$) & 15.33\\
$\sigma_{v}$ & 0.016 & (0.014,0.017) & 20.22\\
$\mu_{v}$ & 9.25$e^{-3}$ & (7.80$e^{-3}$,0.011) & 34.11\\\hline
$\delta_{0}^{p}$ & 0.134 & (0.109,0.173) & 14.07\\
$\alpha_{p}$ & 0.097 & (0.073,0.128) & 9.57\\
$\beta_{pp}$ & 0.062 & (0.048,0.079) & 11.67\\\hline
$\delta_{0}^{v}$ & 0.123 & (0.084,0.165) & 46.73\\
$\alpha_{v}$ & 0.035 & (0.024,0.047) & 83.77\\
$\beta_{vv}$ & 0.031 & (0.021,0.041) & 86.48\\
$\beta_{vp}$ & 5.94$e^{-4}$ & (1.43$e^{-5}$,2.04$e^{-3}$) & 1.75\\\hline
$\Pr\left(  \Delta N_{t}^{v}=1|\Delta N_{t}^{p}=1\right)  $ & 0.104 &
(0.063,0.151) & 24.76\\
$\Pr\left(  \Delta N_{t+1}^{v}=1|\Delta N_{t}^{p}=1\right)  $ & 0.114 &
(0.069,0.163) & 27.59\\\hline\hline
\end{tabular}

\end{center}

\newpage

Table A2: Posterior summaries for model $\mathcal{M}_{2}$, based on S\&P500
stock index data from January 3, 1996 to June 23, 2014, inclusive.

\begin{center}%
\begin{tabular}
[c]{lccc}
&  &  & \\
\multicolumn{4}{l}{$\mathcal{M}_{2}$: $\beta_{vp}=\beta_{vp}^{\left(
-\right)  }=0$}\\\hline\hline
Parameter & MPM & 95\%\ HPD interval & Inefficiency Factor\\\hline\hline
$\mu$ & 0.196 & (0.140,0.252) & 1.52\\
$\gamma$ & -8.674 & (-9.963,-5.655) & 0.98\\
$\rho$ & -0.354 & (-0.419,-0.290) & 6.39\\\hline
$\mu_{p}$ & -0.424 & (-0.442,-0.407) & 7.82\\
$\gamma_{p}$ & 11.655 & (10.564,12.810) & 25.73\\
$\sigma_{p}$ & 0.207 & (0.189,0.224) & 12.99\\
$\pi_{p}$ & 0.383 & (0.301,0.462) & 11.46\\\hline
$\alpha$ & 8.57$e^{-4}$ & (1.73$e^{-5}$,3.28$e^{-3}$) & 1.92\\
$\beta$ & 0.797 & (0.625,0.945) & 17.82\\
$\sigma_{M_{p}}$ & 0.183 & (0.164,0.202) & 12.72\\\hline
$\psi_{0}$ & 1.114 & (0.923,1.316) & 130.50\\
$\psi_{1}$ & 1.316 & (1.280,1.356) & 94.00\\
$\sigma_{BV}$ & 0.436 & (0.421,0.450) & 6.60\\\hline
$\kappa$ & 0.114 & (0.091,0.137) & 53.80\\
$\theta$ & 8.08$e^{-3}$ & (7.31$e^{-3}$,8.93$e^{-3}$) & 13.79\\
$\sigma_{v}$ & 0.016 & (0.014,0.017) & 20.41\\
$\mu_{v}$ & 8.95$e^{-3}$ & (7.55$e^{-3}$,0.011) & 44.31\\\hline
$\delta_{0}^{p}$ & 0.135 & (0.109,0.171) & 15.03\\
$\alpha_{p}$ & 0.097 & (0.072,0.127) & 10.63\\
$\beta_{pp}$ & 0.062 & (0.047,0.080) & 13.97\\\hline
$\delta_{0}^{v}$ & 0.122 & (0.082,0.161) & 43.19\\
$\alpha_{v}$ & 0.032 & (0.019,0.048) & 170.82\\
$\beta_{vv}$ & 0.028 & (0.017,0.042) & 159.08\\\hline
$\Pr\left(  \Delta N_{t}^{v}=1|\Delta N_{t}^{p}=1\right)  $ & 0.108 &
(0.071,0.146) & 16.05\\
$\Pr\left(  \Delta N_{t+1}^{v}=1|\Delta N_{t}^{p}=1\right)  $ & 0.117 &
(0.075,0.157) & 17.94\\\hline\hline
\end{tabular}

\end{center}

\newpage

Table A3: Posterior summaries for model $\mathcal{M}_{3}$, based on S\&P500
stock index data from January 3, 1996 to June 23, 2014, inclusive.

\begin{center}%
\begin{tabular}
[c]{lccc}
&  &  & \\
\multicolumn{4}{l}{$\mathcal{M}_{3}$: $\Delta N_{t}^{p}=\Delta N_{t}^{v}\text{
for all }t=1,...,T$}\\\hline\hline
Parameter & MPM & 95\%\ HPD interval & Inefficiency Factor\\\hline\hline
$\mu$ & 0.232 & (0.166,0.292) & 1.14\\
$\gamma$ & -7.910 & (-9.925,-3.900) & 1.11\\
$\rho$ & -0.397 & (-0.451,-0.340) & 6.66\\\hline
$\mu_{p}$ & -0.436 & (-0.454,-0.419) & 8.74\\
$\gamma_{p}$ & 12.178 & (11.027,13.057) & 36.97\\
$\sigma_{p}$ & 0.197 & (0.177,0.217) & 14.59\\
$\pi_{p}$ & 0.518 & (0.243,0.795) & 170.94\\\hline
$\alpha$ & 0.088 & (0.081,0.096) & 0.99\\
$\beta$ & 0.700 & (0.406,0.923) & 1.02\\
$\sigma_{M_{p}}$ & 0.195 & (0.175,0.214) & 14.02\\\hline
$\psi_{0}$ & 1.245 & (1.052,1.500) & 185.76\\
$\psi_{1}$ & 1.357 & (1.318,1.403) & 129.77\\
$\sigma_{BV}$ & 0.455 & (0.442,0.469) & 3.51\\\hline
$\kappa$ & 0.034 & (0.026,0.043) & 2.39\\
$\theta$ & 0.014 & (0.012,0.015) & 1.17\\
$\sigma_{v}$ & 0.020 & (0.018,0.021) & 22.14\\
$\mu_{v}$ & 8.05$e^{-3}$ & (8.00$e^{-3}$,8.24$e^{-3}$) & 1.03\\\hline
$\delta_{0}^{p}=\delta_{0}^{v}$ & 2.22$e^{-4}$ & (5.41$e^{-6}$,8.35$e^{-4}$) &
1.03\\
$\alpha_{p}=\alpha_{v}$ & 0.140 & (0.044,0.282) & 20.97\\
$\beta_{pp}=\beta_{vv}$ & 0.078 & (4.24$e^{-3}$,0.203) & 10.25\\\hline
$\Pr\left(  \Delta N_{t}^{v}=1|\Delta N_{t}^{p}=1\right)  $ & 1.00 & N/A &
N/A\\
$\Pr\left(  \Delta N_{t+1}^{v}=1|\Delta N_{t}^{p}=1\right)  $ & 0.00 & N/A &
N/A\\\hline\hline
\end{tabular}

\end{center}

\newpage

Table A4: Posterior summaries for model $\mathcal{M}_{4}$, based on S\&P500
stock index data from January 3, 1996 to June 23, 2014, inclusive.

\begin{center}%
\begin{tabular}
[c]{lccc}
&  &  & \\
\multicolumn{4}{l}{$\mathcal{M}_{4}$: $\Delta N_{t}^{v}=0\text{ for all
}t=1,...,T$}\\\hline\hline
Parameter & MPM & 95\%\ HPD interval & Inefficiency Factor\\\hline\hline
$\mu$ & 0.205 & (0.142,0.263) & 1.40\\
$\gamma$ & -8.270 & (-9.946,-4.630) & 1.12\\
$\rho$ & -0.336 & (-0.388,-0.283) & 5.15\\\hline
$\mu_{p}$ & -0.436 & (-0.454,-0.419) & 6.90\\
$\gamma_{p}$ & 12.728 & (11.617,13.921) & 22.18\\
$\sigma_{p}$ & 0.210 & (0.191,0.227) & 12.42\\
$\pi_{p}$ & 0.405 & (0.330,0.484) & 11.91\\\hline
$\alpha$ & 8.51$e^{-4}$ & (2.19$e^{-5}$,3.20$e^{-3}$) & 1.82\\
$\beta$ & 0.800 & (0.602,0.941) & 22.64\\
$\sigma_{M_{p}}$ & 0.182 & (0.162,0.201) & 13.54\\\hline
$\psi_{0}$ & 1.340 & (1.157,1.518) & 130.98\\
$\psi_{1}$ & 1.366 & (1.330,1.402) & 92.92\\
$\sigma_{BV}$ & 0.450 & (0.437,0.464) & 3.85\\\hline
$\kappa$ & 0.036 & (0.028,0.045) & 2.43\\
$\theta$ & 0.013 & (0.012,0.015) & 1.11\\
$\sigma_{v}$ & 0.020 & (0.018,0.021) & 18.91\\\hline
$\delta_{0}^{p}$ & 0.138 & (0.110,0.177) & 17.63\\
$\alpha_{p}$ & 0.097 & (0.072,0.127) & 11.07\\
$\beta_{pp}$ & 0.062 & (0.047,0.078) & 14.67\\\hline
$\Pr\left(  \Delta N_{t}^{v}=1|\Delta N_{t}^{p}=1\right)  $ & 0.00 & N/A &
N/A\\
$\Pr\left(  \Delta N_{t+1}^{v}=1|\Delta N_{t}^{p}=1\right)  $ & 0.00 & N/A &
N/A\\\hline\hline
\end{tabular}

\end{center}

\newpage

Table A5: Posterior summaries for model $\mathcal{M}_{5}$, based on S\&P500
stock index data from January 3, 1996 to June 23, 2014, inclusive.

\begin{center}%
\begin{tabular}
[c]{lccc}
&  &  & \\
\multicolumn{4}{l}{$\mathcal{M}_{5}$: $\left\{
\begin{tabular}
[c]{l}%
$\delta_{t}^{p}=\alpha_{p_{0}}+\alpha_{p}V_{t}\text{ and }$\\
$\delta_{t}^{v}=\alpha_{v_{0}}+\alpha_{v}V_{t}\text{ for all }t=1,...,T$%
\end{tabular}
\ \ \ \ \ \right.  $}\\\hline\hline
Parameter & MPM & 95\%\ HPD interval & Inefficiency Factor\\\hline\hline
$\mu$ & 0.196 & (0.135,0.252) & 1.54\\
$\gamma$ & -8.639 & (-9.960,-5.533) & 1.04\\
$\rho$ & -0.328 & (-0.390,-0.264) & 6.31\\\hline
$\mu_{p}$ & -0.420 & (-0.437,-0.403) & 7.40\\
$\gamma_{p}$ & 11.223 & (10.055,12.249) & 26.18\\
$\sigma_{p}$ & 0.208 & (0.189,0.226) & 13.85\\
$\pi_{p}$ & 0.381 & (0.291,0.467) & 16.51\\\hline
$\alpha$ & 9.24$e^{-4}$ & (2.45$e^{-5}$,3.44$e^{-3}$) & 1.64\\
$\beta$ & 0.770 & (0.563,0.941) & 23.25\\
$\sigma_{M_{p}}$ & 0.183 & (0.163,0.203) & 14.41\\\hline
$\psi_{0}$ & 1.062 & (0.865,1.248) & 152.61\\
$\psi_{1}$ & 1.308 & (1.270,1.346) & 111.28\\
$\sigma_{BV}$ & 0.440 & (0.426,0.453) & 5.07\\\hline
$\kappa$ & 0.087 & (0.073,0.100) & 17.21\\
$\theta$ & 9.13$e^{-3}$ & (8.20$e^{-3}$,0.010) & 11.38\\
$\sigma_{v}$ & 0.016 & (0.015,0.018) & 23.18\\
$\mu_{v}$ & 0.012 & (9.63$e^{-3}$,0.015) & 51.06\\\hline
$\delta_{0}^{p}$ & 0.140 & (0.111,0.189) & 21.36\\
$\alpha_{p}$ & 0.161 & (3.73$e^{-3}$,0.597) & 2.02\\\hline
$\delta_{0}^{v}$ & 0.066 & (0.046,0.086) & 22.23\\
$\alpha_{v}$ & 2.807 & (2.338,2.967) & 33.32\\\hline
$\Pr\left(  \Delta N_{t}^{v}=1|\Delta N_{t}^{p}=1\right)  $ & 0.052 &
(0.032,0.075) & 8.01\\
$\Pr\left(  \Delta N_{t+1}^{v}=1|\Delta N_{t}^{p}=1\right)  $ & 0.056 &
(0.035,0.080) & 8.90\\\hline\hline
\end{tabular}

\end{center}

\newpage

Table A6: Posterior summaries for model $\mathcal{M}_{6}$, based on S\&P500
stock index data from January 3, 1996 to June 23, 2014, inclusive.

\begin{center}
\bigskip%
\begin{tabular}
[c]{lccc}
&  &  & \\
\multicolumn{4}{l}{$\mathcal{M}_{6}$: $\left\{
\begin{tabular}
[c]{l}%
$\delta_{t}^{p}=\alpha_{p_{0}}+\alpha_{p1}V_{t}+\alpha_{p2}V_{t}^{2}\text{ and
}$\\
$\delta_{t}^{p}=\alpha_{v_{0}}+\alpha_{v1}V_{t}+\alpha_{v2}V_{t}^{2}\text{ for
all }t=1,...,T$%
\end{tabular}
\ \ \ \ \ \right.  $}\\\hline\hline
Parameter & MPM & 95\%\ HPD interval & Inefficiency Factor\\\hline\hline
$\mu$ & 0.192 & (0.1304,0.250) & 1.57\\
$\gamma$ & -8.589 & (-9.959,-5.460) & 0.99\\
$\rho$ & -0.324 & (-0.389,-0.259) & 7.67\\\hline
$\mu_{p}$ & -0.419 & (-0.436,-0.403) & 7.68\\
$\gamma_{p}$ & 11.103 & (10.060,12.391) & 32.74\\
$\sigma_{p}$ & 0.209 & (0.187,0.226) & 16.86\\
$\pi_{p}$ & 0.383 & (0.301,0.464) & 12.13\\\hline
$\alpha$ & 9.53$e^{-4}$ & (2.60$e^{-5}$,3.46$e^{-3}$) & 1.88\\
$\beta$ & 0.701 & (0.431,0.923) & 40.62\\
$\sigma_{M_{p}}$ & 0.183 & (0.162,0.205) & 17.80\\\hline
$\psi_{0}$ & 1.037 & (0.863,1.253) & 144.04\\
$\psi_{1}$ & 1.303 & (1.268,1.344) & 103.62\\
$\sigma_{BV}$ & 0.441 & (0.428,0.455) & 5.00\\\hline
$\kappa$ & 0.081 & (0.069,0.095) & 17.70\\
$\theta$ & 9.52$e^{-3}$ & (8.55$e^{-3}$,0.011) & 9.62\\
$\sigma_{v}$ & 0.017 & (0.015,0.018) & 20.60\\
$\mu_{v}$ & 0.013 & (0.010,0.016) & 36.47\\\hline
$\delta_{0}^{p}$ & 0.158 & (0.112,0.247) & 18.45\\
$\alpha_{p1}$ & 0.425 & (9.07$e^{-3}$,1.443) & 6.12\\
$\alpha_{p2}$ & -1.977 & (-9.317,5.539) & 1.21\\\hline
$\delta_{0}^{v}$ & 0.054 & (0.038,0.073) & 18.45\\
$\alpha_{v1}$ & 2.549 & (1.547,2.950) & 10.56\\
$\alpha_{v2}$ & 0.790 & (-6.394,8.867) & 1.34\\\hline
$\Pr\left(  \Delta N_{t}^{v}=1|\Delta N_{t}^{p}=1\right)  $ & 0.046 &
(0.022,0.076) & 24.54\\
$\Pr\left(  \Delta N_{t+1}^{v}=1|\Delta N_{t}^{p}=1\right)  $ & 0.049 &
(0.024,0.078) & 23.05\\\hline\hline
\end{tabular}

\end{center}

\newpage

Table A7: Posterior summaries for model $\mathcal{M}_{7}$, based on S\&P500
stock index data from January 3, 1996 to June 23, 2014, inclusive.

\begin{center}%
\begin{tabular}
[c]{lccc}
&  &  & \\
\multicolumn{4}{l}{$\mathcal{M}_{7}$: $\left\{
\begin{tabular}
[c]{l}%
$\delta_{t}^{p}=\frac{\exp\left(  \alpha_{p_{0}}+\alpha_{p}V_{t}\right)
}{1+\exp\left(  \alpha_{p_{0}}+\alpha_{p}V_{t}\right)  }$ and\\
$\delta_{t}^{v}=\frac{\exp\left(  \alpha_{v_{0}}+\alpha_{v}V_{t}\right)
}{1+\exp\left(  \alpha_{v_{0}}+\alpha_{v}V_{t}\right)  }\text{ for all
}t=1,...,T$%
\end{tabular}
\ \ \ \ \ \right.  $}\\\hline\hline
Parameter & MPM & 95\%\ HPD interval & Inefficiency Factor\\\hline\hline
$\mu$ & 0.202 & (0.143,0.258) & 1.58\\
$\gamma$ & -8.759 & (-9.963,-5.777) & 1.06\\
$\rho$ & -0.322 & (-0.389,-0.254) & 6.41\\\hline
$\mu_{p}$ & -0.419 & (-0.435,-0.402) & 7.91\\
$\gamma_{p}$ & 11.134 & (9.958,12.241) & 33.88\\
$\sigma_{p}$ & 0.209 & (0.191,0.226) & 13.13\\
$\pi_{p}$ & 0.385 & (0.307,0.471) & 11.48\\\hline
$\alpha$ & 9.02$e^{-4}$ & (2.43$e^{-5}$,3.20$e^{-3}$) & 1.76\\
$\beta$ & 0.759 & (0.534,0.944) & 36.90\\
$\sigma_{M_{p}}$ & 0.182 & (0.162,0.201) & 13.95\\\hline
$\psi_{0}$ & 1.024 & (0.827,1.230) & 210.16\\
$\psi_{1}$ & 1.298 & (1.258,1.338) & 151.42\\
$\sigma_{BV}$ & 0.429 & (0.415,0.443) & 6.36\\\hline
$\kappa$ & 0.143 & (0.110,0.161) & 51.51\\
$\theta$ & 8.04$e^{-3}$ & (7.31$e^{-3}$,8.87$e^{-3}$) & 15.21\\
$\sigma_{v}$ & 0.016 & (0.015,0.018) & 22.10\\
$\mu_{v}$ & 8.31$e^{-3}$ & (7.05$e^{-3}$,0.010) & 47.81\\\hline
$\delta_{0}^{p}$ & 0.143 & (0.110,0.200) & 43.62\\
$\alpha_{p_{0}}$ & -1.823 & (-2.108,-1.417) & 40.32\\
$\alpha_{p}$ & 1.171 & (0.035,3.877) & 2.95\\\hline
$\delta_{0}^{v}$ & 0.159 & (0.111,0.209) & 75.80\\
$\alpha_{v_{0}}$ & -3.083 & (-3.430,-2.771) & 34.50\\
$\alpha_{v}$ & 67.697 & (51.550,86.089) & 117.67\\\hline
$\Pr\left(  \Delta N_{t}^{v}=1|\Delta N_{t}^{p}=1\right)  $ & 0.139 &
(0.086,0.194) & 36.05\\
$\Pr\left(  \Delta N_{t+1}^{v}=1|\Delta N_{t}^{p}=1\right)  $ & 0.147 &
(0.091,0.205) & 37.04\\\hline\hline
\end{tabular}

\end{center}

\newpage

Table A8: Posterior summaries for model $\mathcal{M}_{8}$, based on S\&P500
stock index data from January 3, 1996 to June 23, 2014, inclusive.

\begin{center}%
\begin{tabular}
[c]{lccc}
&  &  & \\
\multicolumn{4}{l}{$\mathcal{M}_{8}$: $\delta_{t}^{p}=\delta_{0}^{p}\text{,
}\delta_{t}^{v}=\delta_{0}^{v}\text{ for all }t=1,...,T$}\\\hline\hline
Parameter & MPM & 95\%\ HPD interval & Inefficiency Factor\\\hline\hline
$\mu$ & 0.195 & (0.135,0.250) & 1.59\\
$\gamma$ & -8.543 & (-9.960,-5.331) & 1.09\\
$\rho$ & -0.331 & (-0.392,-0.270) & 6.86\\\hline
$\mu_{p}$ & -0.422 & (-0.442,-0.405) & 11.28\\
$\gamma_{p}$ & 11.397 & (10.135,12.977) & 49.19\\
$\sigma_{p}$ & 0.210 & (0.191,0.227) & 13.29\\
$\pi_{p}$ & 0.385 & (0.304,0.469) & 13.29\\\hline
$\alpha$ & 8.94$e^{-4}$ & (2.28$e^{-5}$,3.28$e^{-3}$) & 1.82\\
$\beta$ & 0.764 & (0.540,0.943) & 24.77\\
$\sigma_{M_{p}}$ & 0.182 & (0.162,0.203) & 15.01\\\hline
$\psi_{0}$ & 1.101 & (0.852,1.374) & 283.17\\
$\psi_{1}$ & 1.318 & (1.269,1.368) & 204.54\\
$\sigma_{BV}$ & 0.447 & (0.433,0.460) & 4.57\\\hline
$\kappa$ & 0.052 & (0.042,0.063) & 14.12\\
$\theta$ & 0.011 & (9.36$e^{-3}$,0.012) & 6.51\\
$\sigma_{v}$ & 0.017 & (0.015,0.018) & 30.89\\
$\mu_{v}$ & 0.017 & (0.012,0.023) & 86.26\\\hline
$\delta_{0}^{p}$ & 0.142 & (0.110,0.197) & 23.80\\\hline
$\delta_{0}^{v}$ & 0.024 & (0.015,0.036) & 18.89\\\hline
$\Pr\left(  \Delta N_{t}^{v}=1|\Delta N_{t}^{p}=1\right)  $ & 0.018 &
(0.007,0.030) & 7.61\\
$\Pr\left(  \Delta N_{t+1}^{v}=1|\Delta N_{t}^{p}=1\right)  $ & 0.019 &
(0.007,0.033) & 10.84\\\hline\hline
\end{tabular}

\end{center}

\bigskip

\newpage Table A9: Posterior summaries for model $\mathcal{M}_{9}$, based on
S\&P500 stock index data from January 3, 1996 to June 23, 2014, inclusive.

\begin{center}%
\begin{tabular}
[c]{lccc}
&  &  & \\
\multicolumn{4}{l}{$\mathcal{M}_{9}$: $\delta_{t}^{p}=0\text{, }\delta_{t}%
^{v}=0\text{ for all }t=1,...,T$}\\\hline\hline
Parameter & MPM & 95\%\ HPD interval & Inefficiency Factor\\\hline\hline
$\mu$ & 0.217 & (0.159,0.272) & 1.18\\
$\gamma$ & -7.916 & (-9.916,-3.922) & 1.21\\
$\rho$ & -0.287 & (-0.342,-0.229) & 4.90\\\hline
$\psi_{0}$ & 0.309 & (0.222,0.404) & 19.02\\
$\psi_{1}$ & 1.090 & (1.071,1.110) & 23.31\\
$\sigma_{BV}$ & 0.470 & (0.457,0.483) & 19.60\\\hline
$\kappa$ & 0.038 & (0.029,0.050) & 1.52\\
$\theta$ & 0.013 & (0.012,0.015) & 1.02\\
$\sigma_{v}$ & 0.023 & (0.021,0.025) & 11.43\\\hline
$\Pr\left(  \Delta N_{t}^{v}=1|\Delta N_{t}^{p}=1\right)  $ & 0.000 & N/A &
N/A\\
$\Pr\left(  \Delta N_{t+1}^{v}=1|\Delta N_{t}^{p}=1\right)  $ & 0.000 & N/A &
N/A\\\hline\hline
\end{tabular}

\bigskip
\end{center}

\newpage Table A10: Posterior summaries for model $\mathcal{M}_{10}$, based on
S\&P500 stock index data from January 3, 1996 to June 23, 2014, inclusive.

\begin{center}%
\begin{tabular}
[c]{lccc}
&  &  & \\
\multicolumn{4}{l}{$\mathcal{M}_{10}$: $\left\{
\begin{tabular}
[c]{l}%
$r_{t}=\sqrt{h_{t}}z_{t}$\\
$h_{t}=\omega+\beta h_{t-1}+\gamma BV_{t-1}$ and\\
$BV_{t}=\xi+\varphi h_{t}+\tau_{1}z_{t}+\tau_{2}\left(  z_{t}^{2}-1\right)
+u_{t}$%
\end{tabular}
\ \ \ \ \ \right.  $}\\\hline\hline
Parameter & MPM & 95\%\ HPD interval & Inefficiency Factor\\\hline
$\omega$ & 0.008 & (0.007,0.009) & 346.80\\
$\beta$ & 0.615 & (0.570,0.647) & 403.06\\
$\gamma$ & 0.083 & (0.067,0.093) & 494.23\\
$\xi$ & -0.079 & (-0.103,-0.068) & 165.17\\
$\varphi$ & 3.901 & (3.454,4.615) & 161.66\\
$\tau_{1}$ & -0.002 & (-0.003,-0.001) & 3.38\\
$\tau_{2}$ & 0.003 & (0.003,0.004) & 3.47\\
$\sigma_{u}$ & 0.034 & (0.033,0.034) & 3.42\\\hline
$\Pr\left(  \Delta N_{t}^{v}=1|\Delta N_{t}^{p}=1\right)  $ & 0.000 & N/A &
N/A\\
$\Pr\left(  \Delta N_{t+1}^{v}=1|\Delta N_{t}^{p}=1\right)  $ & 0.000 & N/A &
N/A\\\hline\hline
\end{tabular}

\bigskip
\end{center}

Table A11: Posterior summaries for model $\mathcal{M}_{11}$, based on S\&P500
stock index data from January 3, 1996 to June 23, 2014, inclusive.

\begin{center}%
\begin{tabular}
[c]{lccc}
&  &  & \\
\multicolumn{4}{l}{$\mathcal{M}_{11}$: $\left\{
\begin{tabular}
[c]{l}%
$r_{t}=\sqrt{h_{t}}z_{t}$\\
$\ln h_{t}=\omega+\beta\ln h_{t-1}+\gamma\ln BV_{t-1}$ and\\
$\ln BV_{t}=\xi+\varphi\ln h_{t}+\tau_{1}z_{t}+\tau_{2}\left(  z_{t}%
^{2}-1\right)  +u_{t}$%
\end{tabular}
\ \ \ \ \ \right.  $}\\\hline\hline
Parameter & MPM & 95\%\ HPD interval & Inefficiency Factor\\\hline
$\omega$ & -0.501 & (-0.587,0.427) & 23.48\\
$\beta$ & 0.527 & (0.492,0.555) & 94.80\\
$\gamma$ & 0.363 & (0.336,0.386) & 78.55\\
$\xi$ & 0.742 & (0.525,0.928) & 11.45\\
$\varphi$ & 1.173 & (1.125,1.214) & 11.72\\
$\tau_{1}$ & -0.094 & (-0.106,-0.084) & 3.41\\
$\tau_{2}$ & 0.059 & (0.054,0.063) & 3.46\\
$\sigma_{u}$ & 0.524 & (0.513,0.533) & 3.38\\\hline
$\Pr\left(  \Delta N_{t}^{v}=1|\Delta N_{t}^{p}=1\right)  $ & 0.000 & N/A &
N/A\\
$\Pr\left(  \Delta N_{t+1}^{v}=1|\Delta N_{t}^{p}=1\right)  $ & 0.000 & N/A &
N/A\\\hline\hline
\end{tabular}

\end{center}